%% file: MGGandMCL.tex
\documentclass{llncs} \usepackage{makeidx} \usepackage{subfigure}

\usepackage{times} \usepackage{epsfig,wrapfig}
\usepackage{amssymb,amsmath,latexsym} \usepackage[english]{babel}
\usepackage{url}  \usepackage{mathabx} \usepackage{ifsym}

\input xy \xyoption{all}

\input{psfig_pp}

\begin{document}

\newcommand{\node}[1]{{\tt#1}}
\newcommand{\figeps}[3]{ 
  \begin{figure}[h!tb]
    \def\epsfsize##1##2{#2##1}%
    \centerline{\epsfbox{Graphics/#1.eps}}
    \caption{#3}
    \label{fig:#1}
  \end{figure}
}

\newcommand{\mor}[1]{\overset{#1}{\longrightarrow}}
\newcommand{\proofend}{\hfill$\Box$}
\newcommand{\abb}[3]{#1 \colon #2 \rightarrow #3}
\newcommand{\pabb}[3]{#1 \colon #2 \PArr #3} \newcommand{\Abb}[2]{(#1
  \rightarrow #2)} \newcommand{\Pabb}[2]{(#1 \PArr #2)}
\newcommand{\morfism}[3]{(#1 \overset{#3} \rightarrow #2)}
\newcommand{\inclusion}[3]{#1 \colon #2 \hookrightarrow #3}
\newcommand{\grule}[5]{#1 \overset{#4}{\longleftarrow} #2
  \overset{#5}{\longrightarrow} #3}
\newcommand{\flattg}{\overline{TG}}
\newcommand{\concretetg}{\widehat{TG}}
\newcommand{\concreterules}[1]{\widehat{#1}}
\newcommand{\unitg}{u_{TG}} \newcommand{\finer}{\leq}
\newcommand{\catgraphs}{\underline{\text{Graph}}}
\newcommand{\catingraphs}{\underline{\text{Inheritance Graph}}}

\newtheorem{defi}{Definition}

\title{Matrix Graph Grammars and Monotone Complex Logics} \author{
  Pedro Pablo P\'erez Velasco, Juan de Lara}


\institute{
  Escuela Polit\'ecnica Superior\\
  Universidad Aut\'onoma de Madrid\\
  \email{\{pedro.perez, juan.delara\}@uam.es} }

\maketitle

\begin{abstract}
  Graph transformation is concerned with the manipulation of graphs by
  means of rules. Graph grammars have been traditionally studied using
  techniques from category theory.  In previous works, we introduced
  Matrix Graph Grammars (MGGs) as a purely algebraic approach for the
  study of graph grammars and graph dynamics, based on the
  representation of graphs by means of their adjacency matrices. MGGs
  have been succesfully applied to problems such as applicability of
  rule sequences, sequentialization and reachability, providing new
  analysis techniques and generalizing and improving previous results.

  Our next objective is to generalize MGGs in order to approach
  computational complexity theory and \emph{static} properties of
  graphs out of the \emph{dynamics} of certain grammars. In the
  present work, we start building bridges between MGGs and complexity
  by introducing what we call \emph{Monotone Complex Logic}, which
  allows establishing a (bijective) link between MGGs and complex
  analysis.  We use this logic to recast the formulation and basic
  building blocks of MGGs as more proper geometric and analytic
  concepts (scalar products, norms, distances). MGG rules can also be
  interpreted -- via operators -- as complex numbers.  Interestingly,
  the subset they define can be characterized as the Sierpinski
  gasket.
\end{abstract}


\section{Introduction}

Graph transformation~\cite{handbook} is concerned with the
manipulation of graphs by means of rules. Similar to Chomsky grammars
for strings, a graph grammar is made of a set of rules, each having a
left and a right hand side (LHS and RHS) graphs and an initial host
graph, to which rules are applied.  The application of a rule to a
host graph is called a derivation step and involves the deletion and
addition of nodes and edges according to the rule specification.
Roughly, when an occurrence of the rule's LHS is found in the graph,
then it can be replaced by the RHS.  Graph transformation has been
successfully applied in many areas of computer science, for example,
to express the valid structure of graphical languages, for the
specification of system behaviour, visual programming, visual
simulation, picture processing and model transformation
(see~\cite{handbook2}). In particular, graph grammars have been used
to specify computations on graphs, as well as to define graph
languages (i.e. sets of graphs with certain properties), thus being
possible to ``translate'' \emph{static} properties of graphs such as
coloring into equivalent properties of dynamical systems (grammars).

In previous work~\cite{MGG_ICGT,MGG_PNGT,MGG_PROLE,MGG_Book} we
developed a new approach to the transformation of \emph{simple}
digraphs. Simple graphs and rules can be represented with Boolean
matrices and vectors and the rewriting can be expressed using Boolean
operators only.  One important point of MGGs is that, as a difference
from other approaches~\cite{handbook}, it explicitly represents the
rule dynamics (addition and deletion of elements), instead of only the
static parts (pre- and post- conditions). Apart from the practical
implications, this fact facilitates new theoretical analysis
techniques such as for example checking independence of a sequence of
arbitrary length and a permutation of it, or obtaining the smallest
graph able to fire a sequence. See~\cite{MGG_Book} for a detailed
account.

In~\cite{MGG_PROLE} we improved our framework with the introduction of
the \emph{nihilation matrix}, which makes explicit some implicit
information in rules: elements that, if present in the host graph,
disable a transformation step. These are all edges not included in the
left hand side, adjacent to nodes deleted by the rule (which would
become dangling) and edges that are added by the production, as in
simple digraphs parallel edges are forbidden. In this paper, we
further develop this idea, as it is natural to consider that a
production transforms pairs of graphs, a ``positive'' one with
elements that must exist (identified by the LHS), and a ``negative''
one, with forbidden elements (identified by the nihilation matrix).

Complexity theory~\cite{Goldreich,Papadimitriou} is concerned with the
study of the {\em intrinsic complexity} of computational tasks.
Traditionally, it has been studied through abstract devices able to
represent the notion of algorithm, such as Turing Machines or Boolean
Circuits~\cite{Vollmer}. Our proposal is to use MGGs instead, as its
algebraic nature allows using results from different branches of
mathematics such as logics, group theory and Boolean algebra.

In this paper we give a first step in the direction of approaching
complexity theory with MGGs, by introducing {\em Monotone Complex
  Logic} (MCL).  Similar to complex numbers, a complex formula in MCL
contains a certainty and a nihil part, both Boolean propositional
formulas. We use MCL terms to encode the ``positive'' part of a simple
digraph (the LHS) and the elements that cannot be found (e.g. the
nihilation matrix).  Using a rational encoding of adjacency matrices,
we can express complex terms referring to simple digraphs into the
unit interval of complex numbers $\mathbb{C}([0,1])$.  Interestingly,
the set of complex numbers defined by valid MCL terms on simple
digraphs is the well-known Sierpinski gasket
fractal~\cite{Mandelbrot}. The rational encoding allows using
geometric and analytic concepts, for example, we have defined a
\textbf{xor}-based norm for MCL terms which can be interpreted as the
number of elementary operations needed to transform one digraph into
another.

Thus, we can use MCL terms to redefine and extend all concepts of
MGGs. In this paper we introduce the encoding of productions in both
its static and dynamic formulations. In the dynamic formulation of a
production, the rule dynamics (element addition and deletion) are also
represented as an MCL term, and thus belong to the Sierpinski gasket
too. We also show the generalization of the main MGG concepts, like
coherence, compatibility, initial digraphs, image of sequences and
G-congruence using MCL.

\noindent \textbf{Paper organization}. Section
\ref{sec:characterization} gives a brief overview of the basic
concepts of MGGs. Section \ref{sec:complexLogics} introduces MCL, used
to establish a link between MGGs and complex analysis.  Section
\ref{sec:numericalRepresentation} encodes graphs as complex numbers,
and a {\em scalar} product, a norm and a notion of distance are
introduced. Section \ref{sec:productionEncoding} encodes productions
as complex numbers and completes the link between MGGs and complex
numbers. Sections \ref{sec:coherenceAndInitialDigraph} and
\ref{sec:compatibilityAndCongruence} generalize the main sequential
results of MGGs such as coherence, compatibility, initial digraphs,
image of sequences and G-congruence. Finally, Sec.
\ref{sec:conclusions} ends with the conclusions and further research.

\section{Matrix Graph Grammars: Basic Concepts}
\label{sec:characterization}

In this section we give a very brief overview of some of the basics of
MGGs, for a detailed account and accesible presentation, the reader is
referred to~\cite{MGG_Book}.

\noindent {\bf Graphs and Rules.} We work with simple digraphs, which
we represent as $(M, V)$ where $M$ is a Boolean matrix for edges (the
graph {\em adjacency} matrix) and $V$ a Boolean vector for vertices or
nodes. We explicitly represent the nodes of the graph with a vector
because rules may add and delete nodes, and thus we mark the existing
nodes with a $1$ in the corresponding position of the vector.
Although nodes and edges can be assigned a type (as
in~\cite{MGG_PROLE}), here we omit it for simplicity.

A production, or rule, $p:L \rightarrow R$ is a partial injective
function of simple digraphs. Using a {\em static formulation}, a rule
is represented by two simple digraphs that encode the left and right
hand sides.

\begin{definition}[Static Formulation of
  Production]\label{def:static_production}
  A production $p:L \rightarrow R$ is statically represented as $p=(
  L=(L^E, L^V); R=(R^E,$ $R^V))$, where $E$ stands for edges and $V$
  for vertices.
\end{definition}

A production adds and deletes nodes and edges; therefore, using a {\em
  dynamic formulation}, we can encode the rule's pre-condition (its
LHS) together with matrices and vectors to represent the addition and
deletion of edges and nodes.

\begin{definition}[Dynamic Formulation of
  Production]\label{def:dynamic_production}
  A production $p:L \rightarrow R$ is dynamically represented as $p=(
  L=(L^E, L^V); e^E, r^E; e^V,$ $r^V)$, where $e^E$ and $e^V$ are the
  deletion Boolean matrix and vector, $r^E$ and $r^V$ are the addition
  Boolean matrix and vector (with a 1 in the position where the
  element is deleted or added respectively).
\end{definition}

The output of rule $p$ is calculated by the Boolean formula $R = p(L)
= r \vee \overline{e} \, L$, which applies to nodes and edges (the
$\wedge$ ({\bf and}) symbol is usually omitted in formulae).

\begin{figure}[htbp]
  \centering \subfigure{
    \includegraphics[scale = 0.4]{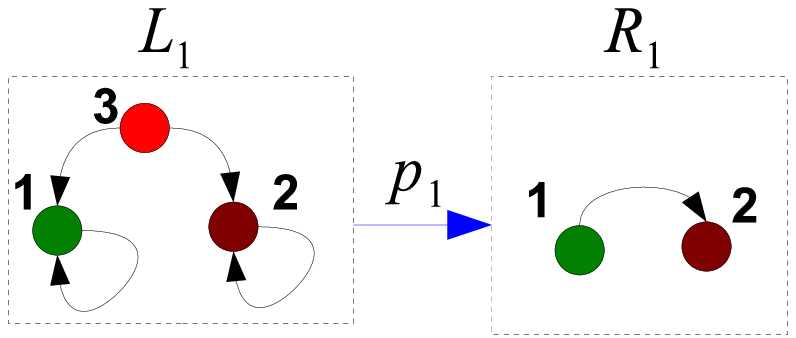}}\hspace{0.3cm}
  \subfigure{
    \includegraphics[scale = 0.7]{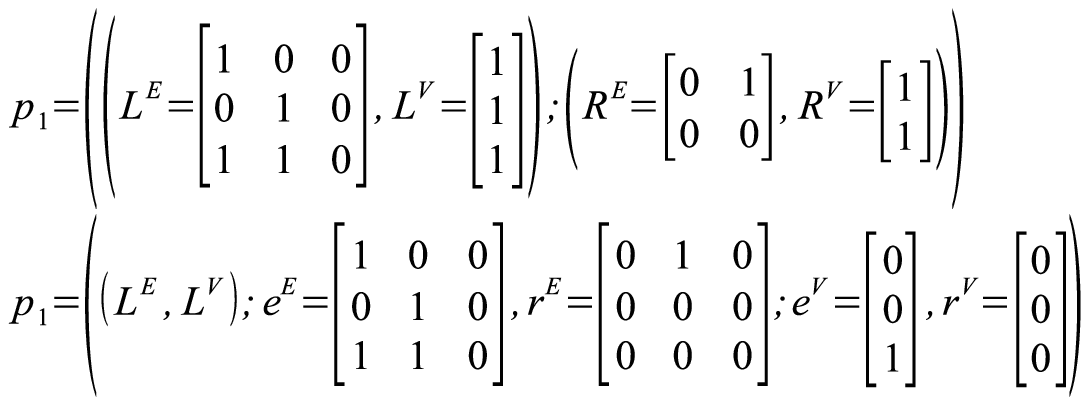}}
  \caption{Simple Production Example (left). Matrix Representation,
    Static and Dynamic (right).}
  \label{fig:example_rule}
\end{figure}

\noindent {\bf Example.} Fig.~\ref{fig:example_rule} shows an example
rule and its associated matrix representation, in its static (right
upper part) and dynamic (right lower part) formulations.$\blacksquare$

In MGGs, we may have to operate graphs of different sizes (i.e.
matrices of different dimensions). An operation called {\em
  completion}~\cite{MGG_ICGT} rearranges rows and columns (so that the
elements that we want to identify match) and inserts zero rows and
columns as needed. For example, if we need to operate with graphs
$L_1$ and $R_1$ in Fig.~\ref{fig:example_rule}, completion adds a
third row and column to $R^E$ (filled with zeros) as well as a third
element (a zero) to vector $R^V$.

\noindent {\bf Compatibility.} A graph $(M,V)$ is compatible if $M$
and $V$ define a simple digraph, i.e. if there are no dangling edges
(edges incident to nodes that are not present in the graph). A rule is
said to be {\em compatible} if its application to a simple digraph
yields a simple digraph (see~\cite{MGG_Book} for the conditions). A
sequence of productions $s_n = p_n; \ldots; p_1$ (where the rule
application order is from right to left) is compatible if the image of
$s_m = p_m; \ldots; p_1$ is compatible, $\forall m \leq n$.

\noindent {\bf Nihilation Matrix.} In order to consider the elements
in the host graph that disable a rule application, rules are extended
with a new graph $K$.\footnote{In \cite{MGG_Book}, $K$ is written
  $N_L$ and $Q$ is written $N_R$. We shall use subindices when dealing
  with sequences in Sec.  \ref{sec:compatibilityAndCongruence}, hence
  the change of notation.  In the definition of production, $L$ stands
  for \emph{left} and $R$ for \emph{right}. The letters that preceed
  them in the alphabet ($K$ and $Q$) have been chosen.} Its associated
matrix specifies the two kinds of forbidden edges: those incident to
nodes deleted by the rule and any edge added by the rule (which cannot
be added twice, since we are dealing with simple
digraphs).\footnote{Nodes are not considered because their addition
  does not generate conflicts of any kind.}


According to the theory developed in \cite{MGG_Book}, no extra effort
is needed from the grammar designer to derive the nihilation matrix,
as $K = p \left(\overline{D} \right)$ with $D = \overline{e^V} \otimes
\overline{e^V}^t$, where $\otimes$ is the tensor product, which sums
up the covariant and contravariant parts and multiplies every element
of the first vector by the whole second vector~\cite{MGG_PROLE}.
Transposition will be represented by ${}^t$.  Please note that given
an arbitrary LHS $L$, a valid nihilation matrix $K$ should satisfy
$L^E K = 0$, that is, the LHS and the nihilation matrix should not
have common edges.

\begin{figure}[htbp]
  \centering \subfigure{
    \includegraphics[scale = 0.4]{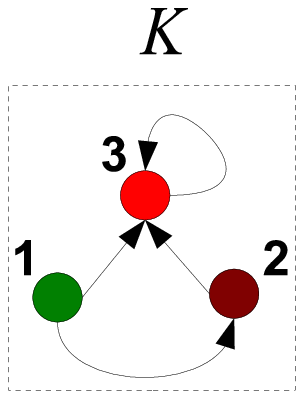}}
  \vline 
  \subfigure{
    \includegraphics[scale = 0.7]{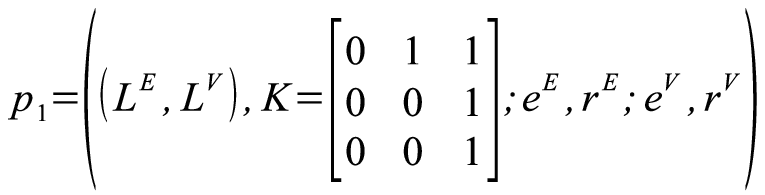}}
  \vline 
  \subfigure{
    \includegraphics[scale = 0.4]{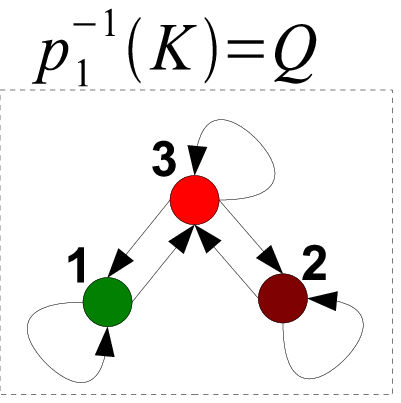}}
  \caption{Nihilation Graph (left). Full Formulation of Production
    (center). Evolution of $K$ (right).}\label{fig:example_nihilation}
\end{figure}

\noindent {\bf Example.} The left of Fig.~\ref{fig:example_nihilation}
shows, in the form of a graph, the nihilation matrix of the rule
depicted in Fig.~\ref{fig:example_rule}. It includes all edges
incident to node $3$ that were not explicitly deleted and all edges
added by $p_1$.  To its right we show the full formulation of $p_1$
which includes the nihilation matrix.$\blacksquare$

As proved in \cite{MGG_Book} (Prop. 7.4.5), the evolution of the
nihilation matrix is fixed by the production. If $R = p(L) = r \vee
\overline{e}L$ then
\begin{equation}
  \label{eq:33}
  Q = p^{-1}(K) = e \vee \overline{r} K,
\end{equation}
being $Q$ the nihilation matrix of the right hand side of the
production $p$. Hence, we have that $(R, Q) = (p(L), p^{-1}(K))$.
Notice that $Q \neq \overline{D}$ in general though it is true that
$\overline{D} \subset Q$. 

\noindent {\bf Example.} The right of
Fig.~\ref{fig:example_nihilation} shows the change in the nihilation
matrix of $p_1$ when the rule is applied. As node $3$ is deleted, no
edge is allowed to stem from it. Self-loops from nodes $1$ and $2$ are
deleted by $p$ so they cannot appear in the resulting
graph.$\blacksquare$

In~\cite{MGG_ICGT} we introduced a functional notation for rules
inspired by the Dirac or bra-ket\footnote{We have followed the
  \emph{mathematical style} instead of the one commonly used in
  physics, which should have been $\left \langle p \, \vert L \right
  \rangle$.} notation~\cite{braket}. Thus, we can depict a rule $p:L
\rightarrow R$ as $R=p(L)=\left\langle L, p\right\rangle$, splitting
the static part (initial state, $L$) from the dynamics (element
addition and deletion, $p$).  Using such formulation, the {\em ket}
operators (i.e. those to the right side of the bra-ket) can be moved
to the {\em bra} (left side) by using their adjoints.  In this work we
recast this notation more properly through MCL.

\noindent {\bf Direct Derivation.} A direct derivation consists on
applying a rule $\abb{p}{L}{R}$ to a graph $G$, through a match
$\abb{m}{L}{G}$ yielding a graph $H$.  In MGGs we use injective
matchings, so given $p:L \rightarrow R$ and a simple digraph $G$ any
$m:L \rightarrow G$ total injective morphism is a match for $p$ in
$G$.  The match is one of the ways of {\em completing} $L$ in $G$.  In
MGGs we do not only consider the elements that should be present in
the host graph $G$ (those in $L$) but also those that should not be
(those in the nihilation matrix, $K$). Hence two morphisms are sought:
$\abb{m_L}{L}{G}$ and $\abb{m_K}{K}{\overline G}$, where $\overline G$
is the complement of $G$, which in the simplest case is just its
negation (see~\cite{MGG_PROLE,MGG_Book}).

\begin{definition}[Direct Derivation]\label{def:directDerivationDef}
  Given rule $p:L \rightarrow R$ and graph $G=(G^E, G^V)$ as in
  Fig.~\ref{fig:matches}(a), $d = \left( p, m \right)$ -- with $m =
  \left( m_L, m_K \right)$ -- is called a direct derivation with
  result $H=p^* \left( G \right)$ if the following conditions are
  fulfilled:
  \begin{enumerate}
  \item There exist $m_L : L \rightarrow G$ and $m_K : K \rightarrow
    \overline{G^E}$ total injective morphisms.
  \item $m_L(n) = m_K(n)$, $\forall n \in L^V$.
  \item The match $m_L$ induces a completion of $L$ in $G$. Matrices
    $e$ and $r$ are then completed in the same way to yield $e^*$ and
    $r^*$. The output graph is calculated as $H=p^*(G)=r^* \vee
    \overline {e^*} G$.
  \end{enumerate}

  \setlength{\unitlength}{0.7cm}
  \begin{figure}[htbp]
    \begin{picture}(15, 4.5)

      \put(9,0)
      {\includegraphics[scale=0.45]{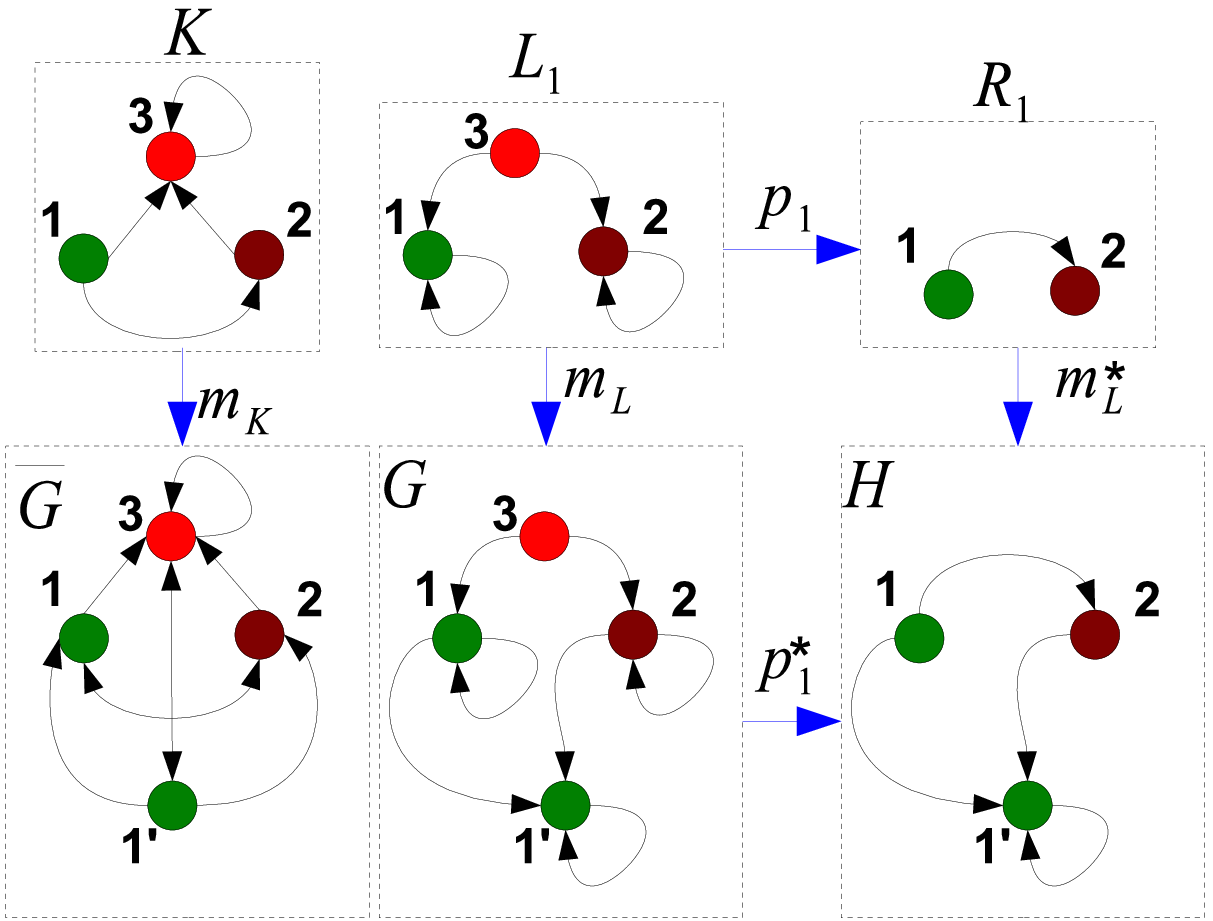}}

      \put(0,4.5) {\xymatrix@C=1cm@R=1cm{
          K\ar[d]_{m_K}           & L\ar[r]^p\ar@{}[rd]|{=}\ar[d]_{m_L} & R\ar@{.>}[d]^{m^*_L} \\
          \overline{G^E}& G\ar@{.>}[r]^{p^*} & H 
        }}

    \end{picture}
    \caption{Direct Derivation (left). Example (right).}
    \label{fig:matches}
  \end{figure}

\end{definition}

\noindent {\bf Remarks.} The square in Fig.~\ref{fig:matches} (a) is a
pushout. Item 2 is needed to ensure that $L$ and $K$ are matched to
the same nodes in $G$.

\noindent {\bf Example} The right of Fig.~\ref{fig:matches} depicts a
direct derivation example using rule $p_1$ shown in
Fig.~\ref{fig:example_rule}, which is applied to a graph $G$ yielding
graph $H$. A morphism from the nihilation matrix to the complement of
$G$, $\abb{m_K}{K}{\overline{G}}$, must also exist for the rule to be
applied.$\blacksquare$

\noindent {\bf Analysis Techniques.}
In~\cite{MGG_ICGT,MGG_PNGT,MGG_PROLE,MGG_Book} we developed some
analysis techniques for MGGs. One of our goals was to analyze rule
sequences independently of a host graph. For its analysis, we {\em
  complete} the sequence by identifying the nodes across rules which
are assummed to be mapped to the same node in the host graph (and thus
rearrange the matrices of the rules in the sequences accordingly).
Once the sequence is completed, our notion of sequence {\em
  coherence}~\cite{MGG_ICGT} allows to know if, for the given
identification, the sequence is potentially applicable (i.e. if no
rule \emph{disturbs} the application of those following it).

Given a completed sequence, the minimal initial digraph (MID) is the
smallest graph that allows applying it. Conversely, the negative
initial digraph (NID) contains all elements that should not be present
in the host graph for the sequence to be applicable. Therefore, the
NID is a graph that should be found in $\overline G$ for the sequence
to be applicable (i.e. none of its edges can be found in $G$).  If the
sequence is not completed (i.e. no overlapping of rules is decided),
we can give the set of all graphs able to fire such sequence or spoil
its application.

Other concepts aim at checking sequential independence (i.e. same
result) between a sequence of rules and a permutation of it. {\em
  G-congruence} detects if two sequences (one permutation of the
other) have the same MID and NID. It returns two matrices and two
vectors, representing two graphs, which are the differences between
the MIDs and NIDs of each sequence respectively. Thus if zero, the
sequences have the same MID and NID. Two coherent and compatible
completed sequences that are G-congruent are sequential independent.

All these concepts have been characterized using operators
$\bigtriangleup$ and $\bigtriangledown$. They extend the structure of
sequence, as explained in \cite{MGG_Book}. Their definition is
included here for future reference:
\begin{eqnarray}
  \label{eq:30}
  \bigtriangleup_{t_0} ^{t_1} \left( F(x,y) \right) & = &
  \bigvee_{y=t_0} ^{t_1} \left( \bigwedge_{x=y} ^{t_1} \left(
      F(x,y) \right) \right) \\
  \bigtriangledown_{t_0} ^{t_1} \left( G(x,y) \right) & = & \bigvee
  _{y=t_0} ^{t_1} \left( \bigwedge_{x=t_0} ^{y} \left( G(x,y)
    \right) \right).
\end{eqnarray}


Productions are the building blocks of sequences and sequences are the
basic construction to study graph dynamics. All these concepts are
further studied and generalized in the present contribution.

Some other important notions such as application conditions, graph
constraints or reachability are just sketched or not even mentioned
and left for further research.

\section{MCL: Monotone Complex Logic}
\label{sec:complexLogics}

In this section we introduce Monotone Complex Logic (MCL), Preliminary
Monotone Complex Algebra (PMCA) and Preliminary Monotone Matrix
Algebra (PMMA). The term ``logic'' in the title should be understood
as in fist-order \emph{logic} or propositional \emph{logic} (arguably,
``calculus'' might be more appropriate). It has been called
\emph{complex} to resemble the similarities with complex numbers and
how they are defined out of the real numbers. \emph{Monotone} because
we are not defining the negation of \emph{complex terms} (see below).

Monotone complex logic is in our opinion of interest by itself, but it
is introduced here due to its usefulness for Matrix Graph Grammars
(MGGs). First, it permits a compact reformulation of grammar rules.
Second, the numerical representation that will be introduced in Def.
\ref{def:rationalEncoding} establishes a link between graphs in MGGs
and $\mathbb{Q}[i]$, although the operations we are interested in are
not addition and multiplication. Also, any production $p$ induces the
evolution of a pair of graphs $(L, K) \overset{p}\longrightarrow (R,
Q) = (p(L), p^{-1}(K))$. Productions will be reinterpreted by encoding
them as complex formulas and representing their actions as a ``Hermite
product''. MCL will allow us to measure the size of graphs via a
natural norm. Finally, sequential notions of MGGs such as
independence, initial digraphs, coherence, etcetera, will be thus
recasted and extended.

\begin{definition}[Complex Formula]\label{def:complexFormula}
  A \emph{complex formula} $z = (a, b)$ consists of a \emph{certainty}
  part '$a$' plus a \emph{nihil} part '$b$', where $a$ and $b$ are
  propositional logic formulas using adjacency matrices as
  propositional variables. Two complex formulas $z_1 = \left(a_1,
    b_1\right)$ and $z_2 = \left(a_2, b_2\right)$ are equal, $z_1 =
  z_2$, if and only if $a_1 = a_2$ and $b_1 = b_2$.
\end{definition}

Monotone Complex Logic is the formal system whose propositional
variables are complex formulas with logical connectives $\{\vee,
\wedge\}$. We will not go further because we are more interested in an
algebraic development of the theory.

Throughout the present contribution, \emph{complex formula},
\emph{complex term} and \emph{Boolean complex} will be used as
synonyms. Next, some basic operations on Boolean complexes are
introduced.

\begin{definition}[Basic Complex Operations]\label{def:basComplexOps}
  Let $z = (a, b)$, $z_1 = (a_1, b_1)$ and $z_2 = (a_2, b_2)$ be
  complex terms. The following operations are defined componentwise:
  \begin{itemize}
  \item Addition: $z_1 \vee z_2 = (a_1 \vee a_2, b_1 \vee b_2)$.
  \item Multiplication: $z_1 \wedge z_2 = z_1 \, z_2 = (a_1 \, a_2
    \vee b_1 \, b_2, a_1 \, b_2 \vee a_2 b_1)$.
  \item Conjugation: $z^* = (\overline{b}, \overline{a})$.
  \item Dot Product: $\left\langle z_1, z_2\right\rangle = z_1 \,
    z_2^*$.
  \end{itemize}
\end{definition}

The notation $\left\langle \cdot, \cdot \right\rangle$ is used for two
reasons. First, we would like to highlight the similarities with
scalar products. There is however no underlying linear space so this
is just a convenient notation. Second, we will see that it coincides
with the functional notation introduced in \cite{MGG_ICGT,MGG_Book}.

The dot product of two Boolean complexes is zero (\emph{orthogonal})
if and only if any element of the first complex term is included in
both the certainty and nihil parts of the second complex term.
Otherwise stated, if $z_1 = (a_1, b_1)$ and $z_2 = (a_2, b_2)$, then
$\left \langle z_1, z_2 \right \rangle = 0 \Leftrightarrow a_1
\overline{a}_2 = a_1 \overline{b}_2 = b_1 \overline{a}_2 = b_1
\overline{b}_2 = 0$. Let's say that $a \prec b$ if $ab = a$, i.e.
whenever $a$ has a 1 $b$ also has a 1 (graph $a$ is \emph{contained}
in graph $b$). Previous identities can be rephrased as $a_1 \prec
a_2$, $a_1 \prec b_2$, $b_1 \prec a_2$ and $b_1 \prec b_2$. This is
equivalent to $(a_1 \vee b_1) \prec (a_2 b_2)$. Orthogonality is
directly related to the common elements of the certainty and nihil
parts.

A particular relevant case is when we consider the dot product of one
element $z = (a, b)$ with itself. In this case we get $(a \vee b)
\prec (ab)$, which is possible if and only if $a = b$. We shall come
back to this issue.

\begin{definition}[Preliminary Monotone Complex Algebra, PMCA]
  \label{def:complexBooleanAlgebra}
  The set $\mathfrak{G}' = \left\{ z \, \vert \, z \right.$ is a
  complex formula$\}$ together with the basic operations introduced in
  Def. \ref{def:basComplexOps} will be known as \emph{preliminary
    monotone complex algebra}.
\end{definition}

We will get rid of the term ``preliminary'' in Def. \ref{def:MCBL},
when not only the adjacency matrix is considered but also the vector
of nodes that make up a simple digraph.


We introduce a subalgebra of the preliminary monotone complex algebra
to be known as \emph{preliminary monotone matrix algebra} (PMMA). It
is useful due to its relationship with MGGs.

\begin{definition}[Preliminary Monotone Matrix Algebra, PMMA]
  \label{def:complexSpace}
  Let $z_i = (a_i, b_i) \in \mathfrak{G}'$. Define the equivalence
  relation $z_1 \sim z_2 \Leftrightarrow \exists c = (c, c) \, \vert
  \, a_2 = a_1 \vee c$ and $b_2 = b_1 \vee c$. Then,
  \begin{equation}
    \label{eq:8}
    \mathfrak{H}' = \mathfrak{G}' / \sim
  \end{equation}
  is the \emph{preliminary monotone matrix algebra}.
\end{definition}

A graph and any of its possible nihilation matrices in MGG do not
share any edge ($L^E K = 0$, refer to \cite{MGG_Book}). So when
representing the left hand side of a production in MCL, its complex
term is made of a digraph in the certainty part and some valid
nihilation matrix in the nihil part. Intuitively, PMMA is made of the
valid complex terms in this sense (i.e. those that do not share any
edge).

It is not difficult to check reflexivity, symmetry and transitivity
for $\sim$.  The equivalence relation permits the simplification of
those elements that appear in both the certainty and nihil parts
(eliminating non-valid complex terms).

A more \emph{complex-analytical} representation can be handy in some
situations and in fact will be preferred for the rest of the present
contribution: $z = (a, b) \longmapsto z = a \vee i \, b$. Define one
element $i$, that we will name \emph{nil term} or \emph{nihil term},
with the property $i \wedge i = 1$, being $i$ itself not equal to 1.
Then, the basic operations of Def. \ref{def:basComplexOps}, following
the same notation, can be rewritten: $z_1 \vee z_2 = \left( a_1 \vee
  a_2 \right) \vee i \left( b_1 \vee b_2 \right)$, $z_1 z_2 = \left(
  a_1 \vee i b_1 \right)\wedge \left(a_2 \vee i b_2 \right)$, $z^* =
\overline{b} \vee i \, \overline{a}$ and the same for the dot product.

Notice that the conjugate of a complex term $z \in \mathfrak{G}'$ that
consists of certainty part only is $z^* = \left( a \vee i0 \right)^* =
1 \vee i \, \overline{a}$.  Similarly for one that consists of nihil
part alone: $z^* = (0 \vee i b)^* = \overline{b} \vee i$. If $z \in
\mathfrak{H}'$ then they further reduce to $a \vee i 0$ and $0 \vee i
b$, respectively, i.e. they are invariant.\footnote{Notice that $1
  \vee i \overline{a} = \left( a \vee \overline{a} \right) \vee i
  \overline{a} = a \vee i 0$ and $\overline{b} \vee i 1 = \overline{b}
  \vee i \left( b \vee \overline{b} \right) = 0 \vee i b$.} Also, the
multiplication reduces to the standard \textbf{and} operation if there
are no nihil parts: $(a_1 \vee i0)(a_2 \vee i0) = a_1 a_2$.

\begin{proposition}\label{def:basicRels}
  Let $x, y, z \in \mathfrak{G}'$ and $z_1, z_2 \in \mathfrak{H}'$.
  Then, $\left \langle x \vee y, z \right\rangle = \left\langle x, z
  \right \rangle \vee \left \langle y, z \right \rangle$, $\left
    \langle z_1, z_2 \right\rangle = \left \langle z_2, z_1 \right
  \rangle^*$ and $\left(z_1 z_2\right)^* = z_1^* z_2^*$.
\end{proposition}

\noindent\emph{Proof}\\
$\square$The first identity is fulfilled by any complex term and
follows directly from the definition. The other two need the
equivalence relation (simplification), i.e. they hold in
$\mathfrak{H}'$ but not necessarily in $\mathfrak{G}'$. For the second
equation just write down the definition of each side of the identity:
\begin{eqnarray}
  \left \langle z_1, z_2 \right \rangle\phantom{^*} & = & \left( a_1
    \overline{b}_2 \vee \overline{a}_2 b_1 \right) \vee i \left(a_1
    \overline{a}_2 \vee b_1 \overline{b}_2 \right) \nonumber \\
  \left \langle z_2, z_1 \right \rangle^* & = & \left[ a_1
    \overline{b}_2 \vee \overline{a}_2 b_1 \vee \left( a_1 b_1 \vee
      \overline{a}_2 \, \overline{b}_2 \right) \right] \vee i \left[
    a_1 \overline{a}_2 \vee b_1 \overline{b}_2 \vee \left( a_1 b_1
      \vee \overline{a}_2 \, \overline{b}_2 \right) \right]. \nonumber
\end{eqnarray}

Terms $a_1 b_1 \vee \overline{a}_2 \overline{b}_2$ vanish as they
appear in both the certainty and nihil parts. The third identity is
proved similarly.$\blacksquare$

Notice however that $\left(z_1 \vee z_2 \right)^* \neq z_1^* \vee
z_2^*$. It can be checked easily as $\left(z_1 \vee z_2 \right)^* =
\left[\left( a_1 \vee a_2 \right) \vee i\left( b_1 \vee b_2
  \right)\right]^* = \overline{b}_1 \overline{b}_2 \vee i\,
\overline{a}_1\overline{a}_2$ but $z_1^* \vee z_2^* =
\left(\overline{b}_1 \vee \overline{b}_2 \right) \vee i\left(
  \overline{a}_1 \vee \overline{a}_2 \right)$. This implies that,
although $\left\langle z_1 \vee z_2 , z \right\rangle = \left\langle
  z_1 , z \right\rangle \vee \left\langle z_2 , z \right\rangle$, we
no longer have \emph{sesquilineality}, i.e. it is not \emph{linear} in
its second component taking into account conjugacy:
\begin{equation}
  \label{eq:13}
  z \left[ \left( \overline{b}_1 \vee \overline{b}_2 \right) \vee i
    \left( \overline{a}_1 \vee \overline{a}_2 \right) \right] =
  \left\langle z, z_1 \vee z_2 \right\rangle \neq \left\langle z, z_1
  \right \rangle \vee \left\langle z, z_2 \right \rangle = z \left[
    \overline{b}_1 \overline{b}_2 \vee i \, \overline{a}_1
    \overline{a}_2 \right]. \nonumber
\end{equation}
In fact the equality $\left\langle z, z_1 \vee z_2 \right\rangle =
\left\langle z, z_1 \right \rangle \vee \left\langle z, z_2 \right
\rangle$ takes place if and only if $z_1 = z_2$.

\section{Numerical Representation, Norm and Distance}
\label{sec:numericalRepresentation}

This section introduces an application $\ell$ that assigns a complex
number in the unit interval $\mathbb{C}\left( [0,1] \right)$ to any
Boolean complex. It is not a homomorphism: neither $\ell(x \vee y) =
\ell (x) + \ell (y)$ nor $\ell(x y) = \ell (x) \ell (y)$ hold.
Application $\ell$ provides some geometric intuition.  A \emph{norm}
and a \emph{conditional norm} are defined out of the dot product of
Def.  \ref{def:basComplexOps}.  Finally we will define the distance
between two complex terms.

\begin{definition}[Rational Enconding]
  \label{def:rationalEncoding}
  Let $g = \left( g^i_j \right)_{i,j \in \{1, \ldots, n\}}$ be a
  simple digraph. Its \emph{rational encoding} is given by
  \begin{equation}
    \ell (g) = \sum_{k = 1}^{n^2}\left( 2^{-k} g^{i_k}_{j_k} \right),
  \end{equation}
  where $i_k = \lceil \frac{k}{n} \rceil$ and $j_ k = k - n \lfloor
  \frac{k}{n} \rfloor$.
\end{definition}

It has become customary to represent the lowest integer above $m$ as
$\lceil m \rceil$ and the biggest integer below $m$ as $\lfloor m
\rfloor$. These functions are known as \emph{ceiling} and
\emph{floor}, respectively. The indices $i_k$ and $j_k$ are just the
integer quotient and the remainder. They are a convenient way to visit
all the elements of the adjacency matrix ordered by columns.

As the elements of $\mathfrak{G}'$ are adjacency matrices we can
define $\mathbb{G}' = \ell \left( \mathfrak{G}' \right)$. If $z =
(a,b) \in \mathfrak{G}'$, then $\ell (z) = \ell(a) + i \ell(b)$.
Analogously, $\mathbb{H}' = \ell \left( \mathfrak{H}' \right)$.

\noindent {\bf Example.} $\ell (L_1)=0'101011_2=0'671875_{10}$, where
$L_1$ is the LHS of $p_1$ in Fig.~\ref{fig:example_rule}. If
$\mathcal{L} = \left[ \begin{array}{cc} 0 & 1 \\ 1 & 0 \end{array}
\right] \vee i \left[ \begin{array}{cc} 1 & 0 \\ 0 & 0 \end{array}
\right]$ then $\ell(\mathcal{L}) = 0'011_2 + i 0'1_2 = 0'375_{10} +
i0'5_{10}$. Subindices indicate the numbering system: base $2$ or base
$10$.$\blacksquare$

The rational encoding is similar to the standard one given in the
literature (which we will call \emph{natural encoding}). The main
difference is that we use negative powers. This, in the limit, would
make us consider the interval $[0,1]$ as the underlying space instead
of the natural numbers $\mathbb{N}$.

\begin{figure}[htbp]
  \centering
  \includegraphics[scale = 0.6]{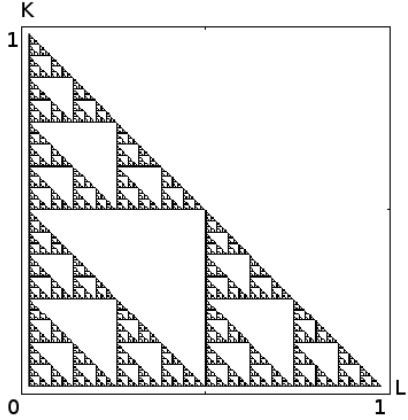}
  \caption{Rational Encoding of the Preliminary Matrix Algebra as a
    Subset of $\mathbb{C}$}
  \label{fig:mgg_CharSet}
\end{figure}

In the present contribution we will deal only with finite graphs.
Hence, the codomain of $\ell$ is $[0,1[ \cap \mathbb{Q}$, in fact only
terminal rationals.\footnote{By hypothesis, binary representation of
  rationals will always be terminating because any dyadic rational
  number $1/2^a$ has a terminating binary numeral (although other
  rational numbers recur). To fix the notation: a terminal number is
  $0'01101$ while a recurring number is $0'\wideparen{011} =
  0'011011011\ldots$} Nonetheless, a small digression seems
appropriate as at some point in the future we will be interested in
the asymptotic behaviour of algorithms and, hence, it will be more
convenient to consider the interval $[0,1]$. Note that $0$ is the
\emph{symmetric} element with respect to the $\vee$ operation.  This
is the graph with no edges. It should be also natural to include 1
because it should be the \emph{neutral} or \emph{identity} element
with respect to the $\wedge$ operation (the graph that has any
possible edge) if the assumption $1 = 0'\wideparen{1}_2$ was made.
However, this assumption is not adequate for MGGs. For example, if we
want to consider the graph $g$ that only has the self-edge $(1,1)$ and
no other one, and the number of nodes of the graph was countable, we
would be tempted to write $g = 0'1_2 \vee 0'0\wideparen{1}_2$. But
$0'0\wideparen{1}_2 = 0'1_2$ with the standard operations in
$\mathbb{C}$. We would thus be asking for the self-edge $(1,1)$ to be
present and not to be present, according to the interpretation of the
certainty and nihil parts in MGG. Seemingly, the Archimedean property
fails as for graphs there are nontrivial infinitesimals. This implies
for example that any diagonal in Fig.  \ref{fig:mgg_CharSet} does not
belong to the MGG characteristic set.

Throughout the present contribution we may make the following abuse of
notation. Let $A$ be the adjacency matrix of some simple digraph. When
we make the operation $A \vee \overline{A}$ we do not obtain $1$ but
$1_A$, which is the matrix with all its elements $1$ according to
vector $A^V$, i.e. $A \vee \overline{A} = A^V \otimes A^V$. This has a
clear relationship with the definition of the nihilation matrix in
\cite{MGG_Book}. See the definition of \emph{conditional norm} below.
Also, we can interpret $1_A$ as the characteristic or indicator
function of all edges potentially incident to nodes of digraph $A$
(the smallest complete digraph that contains $A$). The notation
$\chi_A(x)$ is also standard. It is defined $\chi_A(x) = 1$ if $x \in
A^V \otimes A^V$ and zero otherwise.

The rational encoding in Def. \ref{def:rationalEncoding} establishes
an injection from the preliminary monotone complex algebra
$\mathfrak{G}'$ into the complex numbers,\footnote{Though, as
  commented in this section, just as sets. The morphism is an
  injection if the number of nodes is fixed. For example, $0'1010_2$
  may be the graph with two nodes and edges $(1,1)$ and $(1,2)$ or
  $0'101000000_2$, the graph with three nodes and edges $(1,1)$ and
  $(3,1)$.} where the certainty part becomes the real part, the nihil
part becomes the imaginary part and the nihil term becomes $i =
\sqrt{-1}$. Figure \ref{fig:mgg_CharSet} represents the preliminary
monotone matrix algebra $\mathfrak{H}'$ as a proper subset of
$\mathbb{C}$ which is a well known fractal.

\begin{proposition}\label{prop:mggFractal}
  The characteristic function of the rational encoding of the
  preliminary matrix algebra approaches the Sierpinski gasket as the
  number of nodes increases.
\end{proposition}

\noindent\emph{Proof}\\
$\square$The set corresponds to the zeros of the \textbf{and} function
(coloring zeros in black). It is the Sierpinski gasket due to the
Lucas correspondence theorem \cite{fine} (see also \cite{Weisstein})
which can be used to compute the binomial coefficient $\binom{L}{K}
\mod 2$ with bitwise operations: $\overline{L} \wedge K$. This tells
us that the parity of the function $\binom{L}{K}$ (this is what the
function $\mod 2$ does) is the same as that of $\overline{L} \wedge
K$. In our case $L$ is the abscissa. Its negation just reverts the
order (it is a symmetry) and does not change the shape of the figure.
As commented above in this section, we do not want the diagonals to
belong to the set in the limit.$\blacksquare$

The dot product introduced in Sec. \ref{sec:complexLogics} induces a
``norm'' $\left\lVert \, \cdot \, \right\rVert$ in $\mathfrak{G}'$.
It is not a norm in the sense of linear algebra (no underlying vector
space) but shares some of its properties and will play a similar role:
it will be used as a means to measure the size of a graph.

\begin{definition}[Norm -- Conditional Norm]\label{def:norm}
  Let $y, z \in \mathfrak{G}'$. Its \emph{norm} is the application
  $\left \lVert \cdot \right \rVert : \mathfrak{G}' \rightarrow [0,1]$
  defined by
  \begin{equation}
    \label{eq:14}
    \left \lVert z \right \rVert = \ell \left( \left\langle z, z
      \right\rangle \right).
  \end{equation}
  The \emph{conditional norm} of $z$ with respect to $y$ is given by
  \begin{equation}
    \label{eq:1}
    \left \lVert z \vert y \right \rVert = \left \lVert z \right
    \rVert_y = \frac{\left \lVert z \wedge y \right \rVert}{\left
        \lVert y \right \rVert}.
  \end{equation}
\end{definition}

The following identities show that the norm (before applying $\ell$)
returns what one would expect. Equation \eqref{eq:99} is particularly
relevant as it states that in MCL the certainty and nihil parts are in
some sense mutually exclusive, which together with eq. \eqref{eq:7}
suggest the definition of $\mathfrak{H}'$ as introduced in Sec.
\ref{sec:complexLogics}. Notice that this fits perfectly well with the
interpretation of $L$ and $K$ given in \cite{MGG_Book}.
\begin{eqnarray}
  \label{eq:6}
  \left\langle (a, 0), (a, 0) \right\rangle & = & \left( a, 0 \right)
  \left( 1, \overline{a}\right) = \left( a \, 1 \vee 0 \,
    \overline{a}, a \, \overline{a} \vee 0 \, 1 \right) = a \\
  \label{eq:98}
  \left\langle (0, b), (0, b) \right\rangle & = & \left( 0, b \right)
  \left( \overline{b}, 1 \right) = \left( b \, 1 \vee 0 \,
    \overline{b}, b \, \overline{b} \vee 0 \, 1 \right) = b \\
  \label{eq:99}
  \left\langle (c, c), (c, c) \right\rangle & = & \left( c, c \right)
  \left( \overline{c}, \overline{c}\right) = \left( c \, \overline{c}
    \vee c \, \overline{c}, c \, \overline{c} \vee c \, \overline{c}
  \right) = 0.
\end{eqnarray}

The dot product of one element with itself gives rise to the following
useful identity:
\begin{equation}
  \label{eq:7}
  \left\langle z, z \right\rangle = z \, z^* = \left( a \overline b
    \vee \overline{a} b \right) \vee i \left( b \overline b \vee a
    \overline{a} \right) = a \oplus b.
\end{equation}

Equation \eqref{eq:7} admits two readings. In first place, it tells
how to factorize one of the basic Boolean operations, \textbf{xor}.
Secondly, and more relevant to us, it justifies the use of
\textbf{xor} as a norm for complex terms.\footnote{Recall that in
  complex analysis we have $z \, z^* = \vert z \vert^2$.} Besides, it
follows directly from the definition that $\left \lVert i \right\rVert
= 1$ and $\left \lVert z^* \right \rVert = \left \lVert z \right
\rVert$.

The conditional norm $\left \lVert \, \cdot \, \vert \, y \, \right
\rVert$ \emph{reduces} the total space to graph $y$. If $z = a_1 \vee
i b_1 \in \mathfrak{G}'$ and $y = a_2 \vee i b_2 \in \mathfrak{G}'$,
after some simple algebraic operations
\begin{equation}
  \label{eq:2}
  \left \lVert z \vert y \right \rVert = \frac{\ell \left( \left( a_1
        \oplus b_1 \right) \left( a_2 \oplus b_2 \right) \right)}{\ell
    \left( a_2 \oplus b_2 \right)}
\end{equation}
is obtained. If we further have that $y, z \in \mathfrak{H}'$, then
\begin{equation}
  \label{eq:3}
  \left \lVert z \vert y \right \rVert = \frac{\ell \left( \left( a_1
        \vee b_1 \right) \left( a_2 \vee b_2 \right) \right)}{\ell
    \left( a_2 \vee b_2 \right)}.
\end{equation}
In particular this implies that $\left \lVert z \vert y \right \rVert
= 1 \Leftrightarrow y$ is a subgraph of $z$. The following proposition
highlights the similarities between norms in linear spaces and the one
introduced in Def. \ref{def:norm}.

\begin{proposition}\label{prop:normProps}
  Let $y, z \in \mathfrak{H}'$ and $z_1, z_2 \in \mathfrak{G}'$. Then,
  for norms and conditional norms we have that
  \begin{eqnarray}
    \label{eq:15}\left \lVert z \right \rVert & = & 0 \Leftrightarrow
    z = 0, \\
    \label{eq:101}\left \lVert y z \right \rVert & = & \left
      \lVert y \right \rVert \wedge \left \lVert z \right \rVert, \\
    \label{eq:100}\left \lVert z_1 \vee z_2 \right \rVert & \leq &
    \left \lVert z_1 \right \rVert \vee \left \lVert z_2 \right
    \rVert.
  \end{eqnarray}
\end{proposition}
\noindent\emph{Proof}\\
$\square$Identity \eqref{eq:15} is derived from \eqref{eq:7}. Some
simple manipulations prove \eqref{eq:101} (where \textbf{and} is
performed bitwise). Inequality \eqref{eq:100} is not difficult either
where, again, \textbf{or} on the right hand side is applied bitwise:
\begin{eqnarray}
  \label{eq:18}
  \left \lVert z_1 \right \rVert \vee \left \lVert z_2 \right \rVert &
  = & \ell ( a_1 \overline{b}_1 ) \vee \ell ( a_2 \overline{b}_2 )
  \vee \ell ( \overline{a}_1 b_1 ) \vee \ell ( \overline{a}_2 b_2 )
  \nonumber \\
  \label{eq:181}
  \left \lVert z_1 \vee z_2 \right \rVert & = & \ell ( a_1
  \overline{b}_1 \overline{b}_2 ) \vee \ell ( a_2 \overline{b}_1 
  \overline{b}_2 ) \vee \ell ( b_1 \overline{a}_1 \overline{a}_2
  ) \vee \ell ( b_2 \overline{a}_1 \overline{a}_2 ). \nonumber
\end{eqnarray}
Comparing term by term, the inequality follows. For example, there
must be at least the same numbers of 1's in $a_1 \overline{b}_1$ than
in $a_1 \overline{b}_1 \overline{b}_2$, so $\ell(a_1 \overline{b}_1)
\geq \ell (a_1 \overline{b}_1 \overline{b}_2)$.$\blacksquare$

The \textbf{xor} operation $z = z_1 \oplus z_2$ is the number of
distinct elements in $z_1$ and $z_2$. The number of ones that appear
in $z$ tells the number of atomic operations that must be performed in
order to transform $z_1$ in $z_2$ (or viceversa) in the sense of MGG
productions. Therefore, it seems natural to define the distance
between two complex terms as follows:

\begin{definition}\label{def:distance}
  Let $z_j = a_j \vee i b_j$ with $z_j \in \mathfrak{G}'$, $j = 1, 2$,
  and define $w_1 = a_1 \vee i a_2$ and $w_2 = b_1 \vee i b_2$. Then,
  the distance between $z_1$ and $z_2$ is given by
  \begin{equation}
    \label{eq:22}
    d(z_1,z_2) = \left \lVert \, w_1 \right \rVert \oplus \left \lVert
      \, w_2 \right \rVert. 
  \end{equation}
\end{definition}

It is an easy exercise to check that $d$ fulfills the axioms of a
metric: $d(x,y) \geq 0$, $d(x,y) = 0 \Leftrightarrow x = y$, $d(x,y) =
d(y,x)$ and the triangle inequality $d(x,z) \leq d(x,y) + d(y,z)$. The
triangle inequality follows from $d(x,z) = d(x,y) \oplus d(y,z)$ and
the fact that $\forall a > 0$, $\forall b > 0$ we have that $a \oplus
b \leq a + b$. See \cite{KaD} for an application of the \textbf{xor}
metric.

Notice that in $\mathfrak{H}'$ only one of the terms contributes to
the norm if $z_2 = p(z_1)$ for some production because the actions on
the certainty part completely determine the nihil part, as proved in
\cite{MGG_Book} (Prop. 7.4.5).

\section{Production Encoding}
\label{sec:productionEncoding}

In this section we introduce the Monotone Complex Algebra and the
Monotone Matrix Algebra, that not only consider edges but also nodes.
Compatibility issues may appear so we study compatibility for a simple
digraph and also for a single production (compatibility for sequences
will be addressed in Sec. \ref{sec:compatibilityAndCongruence}). Next
we turn to one of the main topics in this paper: how to characterize
MGG productions using MCL and the dot product of Def.
\ref{def:basComplexOps}. The section ends introducing \emph{swaps} and
providing some geometric interpretations.

To get rid of the ``preliminary'' term in the definitions of
$\mathfrak{G}'$ and $\mathfrak{H}'$ (Defs.
\ref{def:complexBooleanAlgebra} and \ref{def:complexSpace}, resp.) we
shall consider an element as being composed of a matrix term and a
vector of nodes. Hence, we have that $\mathcal{L} = \left( L^E \vee i
  K^E, L^V \vee i K^V \right)$ where $E$ stands for \emph{edge} and
$V$ for \emph{vertex}.\footnote{If an equation is applied to both
  edges and nodes then the superindices will be omitted. They will
  also be omitted if it is clear from context which one we refer to.}
Notice that $L^E \vee i K^E$ are matrices and $L^V \vee i K^V$ are
vectors.

\begin{definition}[Monotone Complex and Matrix Algebras]
  \label{def:MCBL}
  $\mathfrak{G}$ and $\mathfrak{H}$ (the \emph{Monotone Complex} and
  \emph{Matrix Algebras}, resp.) are defined as their preliminary
  counterparts -- see Defs. \ref{def:complexBooleanAlgebra} and
  \ref{def:complexSpace} -- but considering elements of the form:
  \begin{equation}
    \label{eq:201}
    \mathcal{L} = \left( L^E \vee i K^E, L^V \vee i K^V \right).
  \end{equation}
  We also introduce $\mathbb{G} = \ell \left( \mathfrak{G} \right)$
  and $\mathbb{H} = \ell \left( \mathfrak{H} \right)$.
\end{definition}

Concerning $\mathfrak{G}$, a production $p:\mathfrak{G} \rightarrow
\mathfrak{G}$ consists of two independent productions $p = (p_1, p_2)$
-- being $p_i$ MGG productions as those introduced in \cite{MGG_Book}
-- one acting on the certainty part and the other on the nihil part:
\begin{equation}
  \label{eq:31}
  \mathcal{R} = p(\mathcal{L}) = p_C(L) \vee i p_N(K) = R \vee i Q.
\end{equation}
As there are no restrictions on $p_C$ and $p_N$ if we stick to
$\mathfrak{G}$, it is true that $\forall g_1, g_2 \in \mathfrak{G}$,
$\exists p$ such that $p(g_1) = g_2$. Recall from Sec.
\ref{sec:characterization} that productions in MGG have $\mathfrak{H}$
as domain and codomain.  Moreover, they must fulfill $p_N = p_C^{-1}$.
Unless otherwise stated, we will concentrate on MGG productions for
the rest of the paper.

We want $p_N$ to be a production so we must split it into two parts:
the one that acts on edges and the one that acts on vertices.
Otherwise there would probably be dangling edges in the nihil part as
soon as the production acts on nodes. The point is that the image of
the nihil part with the operations specified by productions are not
graphs in general, unless we restrict to edges and keep nodes apart.
This behaviour is unimportant and should not be misleading.

\begin{figure}[htbp]
  \centering
  \includegraphics[scale = 0.58]{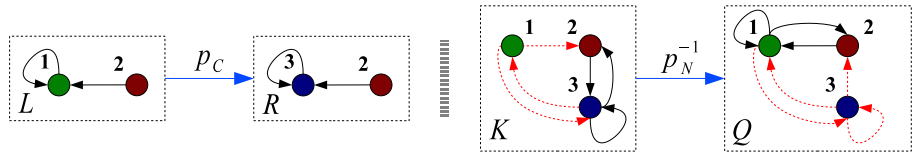}
  \caption{Potential Dangling Edges in the Nihilation Part}
  \label{fig:ex4}
\end{figure}

\noindent \textbf{Example}.$\square$To the left of Fig. \ref{fig:ex4}
we have drawn the certainty part of a production $p$ that deletes node
1 (along with two incident edges) and adds node 3 (and two incident
edges). Its nihil counterpart for edges is depicted to the right of
the same figure. Notice that node $1$ should not be included in $K$
because it appears in $L$ and we would be simultaneously demanding its
presence and its absence. Therefore, edges $(1,3)$, $(2,1)$ and
$(3,1)$ -- those with a red dotted line -- would be dangling in $K$
(red dotted edges do belong to the graphs they appear on). The same
reasoning shows that something similar happens in $Q$ but this time
with edges $(1,3)$, $(3,1)$, $(3,2)$ and $(3,3)$ and node 3.

This is the reason to consider nodes and edges independently in the
nihil parts of graphs and productions. In $K$, as nodes $1$ and $3$
belong to $L$, it should not make much sense to include them in $K$
too, for if $K$ dealt with nodes we would be demanding their presence
and their abscense. In $Q$ the production adds node $3$ and something
similar happens.$\blacksquare$

Now that nodes are considered compatibility issues in the certainty
part may show up. The determination of compatibility for a simple
digraph under MCL is almost straightforward. Let $g = (g^E_C \vee i
g^E_N, g^V_C \vee i g^V_N) \in \mathfrak{H}$. Potential dangling edges
are given by $\overline{D}_g = \overline{\overline{g^V_C} \otimes
  \overline{g^V_C}}$, so the graph $g$ will be compatible if $g^E_C \,
\overline{D}_g = 0$. If $g \in \mathfrak{H}$ there are no common
elements between the certainty and nihil parts and $\overline{D}_g
\prec g^E_N$.

A production $p(\mathcal{L}) = p(L \vee iK) = R \vee i Q =
\mathcal{R}$ is compatible if it preserves compatibility, i.e. if it
transforms a compatible digraph into a compatible digraph. This
amounts to saying that $RQ = 0$.

Recall from Sec. \ref{sec:characterization} that grammar rules actions
are specified through \emph{erasing} and \emph{addition} matrices, $e$
and $r$ respectively. Because $e$ acts on elements that must be
present and $r$ on those that should not exist, it seems natural to
encode a production as
\begin{equation}
  \label{eq:25}
  p = e \vee i r.
\end{equation}

Our next objective is to use the dot product -- see Def.
\ref{def:basComplexOps} -- to represent the application of a
production. This way, a unified approach would be obtained. To this
end define the operator $P: \mathfrak{G} \rightarrow \mathfrak{G}$ by
\begin{equation}
  \label{eq:26}
  p = e \vee i r \longmapsto P(p) = \overline{e} \, \overline{r} \vee i
  \left( e \vee r \right).
\end{equation}

\begin{proposition}[Production]\label{prop:production}
  Let $\mathcal{L}$ and $\mathcal{R}$ be the left and right hand
  sides, resp., as in Def. \ref{def:MCBL}, eq. (\ref{eq:201}), and $P$
  as defined in eq. \eqref{eq:26}. Then,
  \begin{equation}
    \label{eq:5}
    \mathcal{R} = \left\langle \mathcal{L}, P(p) \right\rangle.
  \end{equation}
\end{proposition}

\noindent\emph{Proof}\\
$\square$The proof is a short exercise that makes use of some
identities which are detailed right afterwards:
\begin{eqnarray}
  \label{eq:16}
  \left\langle \mathcal{L}, P(p) \right\rangle & = & \left\langle
    \left( L, K \right), \left(\overline{e} \, \overline{r}, e \vee
      r\right)  \right\rangle = \nonumber \\
  & = & \left(\overline{e} \, \overline{r} L \vee (e \vee r)K ,
    \overline{e}\, \overline{r}K \vee (e \vee r)L \right) = \nonumber
  \\
  & = & \left(r \vee \overline{e} L, e \vee \overline{r} K \right)
  = \left(p(L), p^{-1}(K) \right) = \mathcal{R}.
\end{eqnarray}
Apart from equation (4.13) of Prop. 4.1.4 in \cite{MGG_Book} which
states that $\overline{r}L = L$, we have used the following
identities:
\begin{eqnarray*}
  (e \vee r)K & = & eK \vee rK = rK = r(r \vee \overline{e}
  \overline{D}) = r. \\
  \overline{e}\, \overline{r}\, K & = & \overline{r} \left(
    \overline{e}\,r \vee \overline{e} \, \overline{e} \overline{D}
  \right) = \overline{r}\, K. \\
  (e \vee r)L & = & eL \vee rL = eL = e.
\end{eqnarray*}
We have also used that $r \overline{e} = r$ (again Prop. 4.1.4 in
\cite{MGG_Book}), $r \overline{D} = r$ due to compatibility and $rL =
0$ almost by definition. Besides, Prop. 7.4.5 in \cite{MGG_Book} has
also been used, which proves that $Q = p^{-1} \left( K
\right)$.$\blacksquare$

\begin{figure}[htbp]
  \centering
  \includegraphics[scale = 0.75]{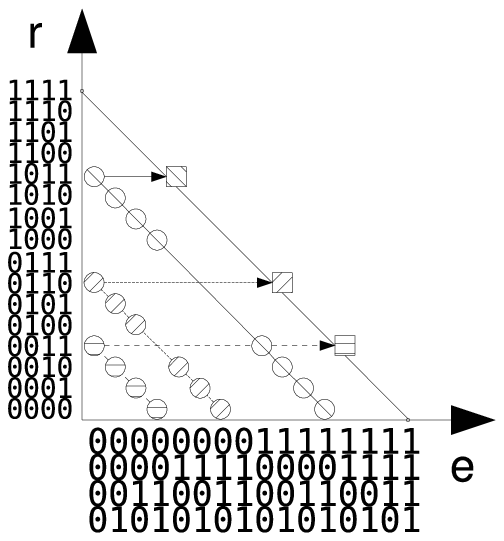} \qquad \qquad
  \includegraphics[scale = 0.77]{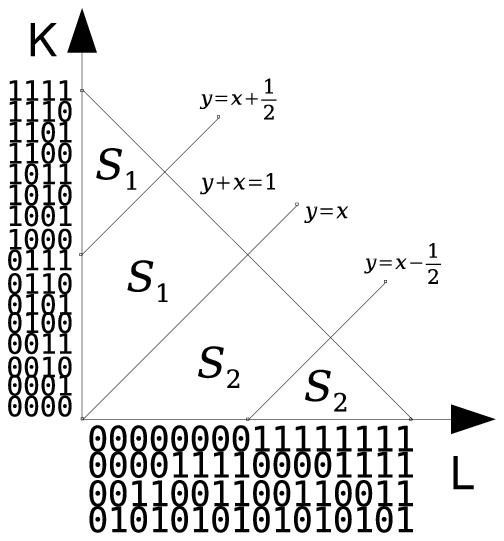}
  \caption{Rational Encoding of the Transformation of Productions via
    Operator $P$}
  \label{fig:imageP}
\end{figure}

The production is defined through operator $P$ instead of directly as
$p = \overline{e} \, \overline{r} \vee i(e \vee r)$ for several
reasons. First, eq. \eqref{eq:25} and its interpretation seem more
natural. Second, $P(p)$ is self-adjoint, i.e.  $P(p)^* = P(p)$, which
in particular implies that $\left \lVert P(p) \vert z \right \rVert =
1$, $\forall p$, being $z$ the ``total'' graph, i.e. the graph with
respect to which \emph{completions} are performed (for
\emph{completion} refer to \cite{MGG_Book}, Sec. 4.2).  Therefore, the
norm would not measure the size of productions (interpreted as graphs
according to eq. \eqref{eq:25}) and we would be forced to introduce a
new norm. This is because
\begin{equation}
  \label{eq:4}
  \left \langle P(p), P(p) \right \rangle = \left( \overline{e}
    \, \overline{r} \vee i (e \vee r) \right) \left( \overline{e} \,
    \overline{r} \vee i (e \vee r) \right)^* = \overline{e} \,
  \overline{r} \vee e \vee r = 1_z. \nonumber
\end{equation}
By way of contrast, $\left \lVert p \right \rVert = e \oplus r = e
\vee r$. With operator $P$ the size of a production is the number of
changes it specifies, which is appropriate for MGGs.\footnote{One of
  our objectives is to look for an appropriate measure of the number
  of actions that would eventually transform one graph into another.}
Moreover, due to eq.  (\ref{eq:25}), geometrically, grammar rules are
also the Sierpinski gasket. In fact, the codomain of operator $P$ is
the diagonal $e + r = 1$ (refer to Figs. \ref{fig:mgg_CharSet} and
\ref{fig:imageP}).

Complex logic encoding puts into a single expression the application
of a grammar rule, both $L$ and $K$. Also, it links the functional
notation introduced in \cite{MGG_Book} and the dot product of Sec.
\ref{sec:complexLogics}.

\begin{theorem}[Surjective Morphism]\label{th:surjMorph}
  There exists a surjective morphism from the set of MGG productions
  on to the set of self-adjoint graphs in $\mathfrak{H}$.
\end{theorem}
\noindent\emph{Proof}\\
$\square$It is not difficult to check that $z$ is self-adjoint if and
only if $\left \lVert z \right \rVert = 1_z$: on the one hand, if $z =
a \vee i \overline{a}$ then $\left \langle z, z \right \rangle = z z^*
= (a \vee i \overline{a})(a \vee i \overline{a}) = a \vee \overline{a}
= 1_z$. On the other hand, if we have $z = a \vee i b$ and $\left
  \lVert z \right \rVert = a \oplus b = 1_z$ then $a = \overline{b}$.
Note that $\left \lVert z \right \rVert = 1_z$ is equivalent to asking
for the conditional norm to be equal to 1 with respect to\footnote{The
  tensor (Kronecker) product in this contribution will always be used
  on nodes. The $V$ superindex will be omitted in this case: $z
  \otimes z^t \equiv z^V \otimes \left( z^t \right)^V$, where ${}^t$
  stands for transposition.} $Z = z \otimes z^t$:
\begin{equation}
  \label{eq:9}
  \left \lVert z \vert Z \right \rVert = \frac{\left \lVert z Z \right
    \rVert}{\left \lVert Z \right \rVert} = \frac{\left \lVert z
    \right \rVert}{\left \lVert Z \right \rVert} = 1. \nonumber
\end{equation}

The surjective morphism is given by operator $P$. Clearly, $P$ is
well-defined for any production. To see that it is surjective, fix
some graph $g = g_1 \vee i g_2$ such that $\left \lVert g \right
\rVert = 1_g$. Then, $g = g_1 \vee i \overline{g}_1$. Any partition of
$\overline{g}_1$ as \textbf{or} of two disjoint digraphs would do.
Recall that productions (as graphs) have the property that their
certainty and nihil parts must be disjoint.$\blacksquare$

The operator $P$ is surjective but not necessarily injective. It
defines an equivalence relation and the corresponding quotient space.
We will be led to a reinterpretation of the notion of production in
Matrix Graph Grammars.

\begin{definition}[Swap]\label{def:swap}
  The swap space is defined as $\mathcal{W} = \mathfrak{H} /
  P(\mathfrak{H})$. An equivalence class in the swap space will be
  called a \emph{swap}. The swap $w$ associated to production $p:
  \mathfrak{H} \rightarrow \mathfrak{H}$ is $w = w_p = P(p)$, i.e. $p
  \in \mathfrak{H} \longmapsto w_p \in \mathcal{W}$.\footnote{Acording
    to eq. \eqref{eq:25}, any element in $\mathfrak{H}$ can be
    interpreted as a production and viceversa.}
\end{definition}

\begin{figure}[htbp]
  \centering
  \includegraphics[scale = 0.45]{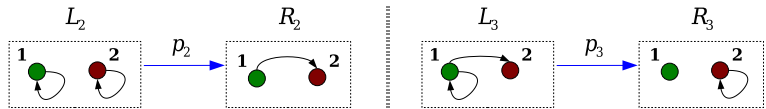}
  \caption{Example of Productions}
  \label{fig:exProds}
\end{figure}

\noindent \textbf{Example}.$\square$Let $p_2$ and $p_3$ be two
productions as those depicted in Fig. \ref{fig:exProds}. Their images
in $\mathcal{H}$ are:
\begin{equation}
  \label{eq:10}
  P(p_2) =
  P(p_3) = \left[ \begin{array}{cc} 0 & 0 \\ 1 & 0 \end{array} \right]
  \vee i \left[ \begin{array}{cc} 1 & 1 \\ 0 & 1 \end{array}
  \right] = w.
\end{equation}

They appear to be very different if we look at their defining matrices
$L_2, L_3$ and $R_2, R_3$ or at their graph representation. Also, they
seem to differ if we look at their erasing and addition matrices:
\begin{equation}
  \label{eq:11}
  e_2 = \left[ \begin{array}{cc} 1 & 0 \\ 0 & 1 \end{array} \right] \quad
  e_3 = \left[ \begin{array}{cc} 1 & 1 \\ 0 & 0 \end{array} \right] \quad
  r_2 = \left[ \begin{array}{cc} 0 & 1 \\ 0 & 0 \end{array} \right] \quad
  r_3 = \left[ \begin{array}{cc} 0 & 0 \\ 0 & 1 \end{array} \right].
  \nonumber
\end{equation}

However, they are the same swap as eq. (\ref{eq:10}) shows, i.e. they
belong to the same equivalence class. Notice that both productions act
on edges $(1,1)$, $(2,2)$ and $(1,2)$ and none of them touches edge
$(2,1)$. This is precisely what eq. \eqref{eq:10} says as we will
promptly see.

Swaps will be of help in studying and classifying the productions of a
grammar. For example, there are 16 different simple digraphs with 2
nodes.  Hence, there are 256 different productions that can be
defined.  However, there are only 16 different swaps. From the point
of view of the number of edges that can be modified, there is 1 swap
that does not act on any element (which includes 16 productions), 4
swaps that act on 1 element, 6 swaps that act on 2 elements, 4 swaps
that act on 3 elements and 1 swap that acts on all
elements.$\blacksquare$

There is a simple geometrical interpretation of $P$ in terms of the
rational encoding of the productions $p$ and the way they are
transformed. Let the principal diagonal be the closest
line\footnote{``Closest'' because the line $x + y = 1$ is approached
  as the number of nodes in the graphs tend to $\infty$.}  to $x + y =
1$. Operator $P$ assigns the same element $P(p)$ in the principal
diagonal to any element $p$ of a parallel line (to the left of Fig.
\ref{fig:imageP} three parallel diagonals are represented ). $P(p)$
has the same certainty part as the biggest certainty part of any $p$
in the parallel line. See Fig. \ref{fig:imageP} for the transformation
of three sets of productions (circles) into their associated swaps
(squares). Geometrically, $P$ can be thought of as the composition of
two projections: one along the corresponding diagonal and another
parallel to the abscissa axis.

The name swap has been chosen because of the way they act on elements
in $\mathfrak{H}$. Following previous example, let's consider the swap
$w = 0'0100 \vee i 0'1011$. Let's also consider a generic element
$\mathcal{L} = 0'L^1_1L^2_1L^1_2L^2_2 \vee i 0'K^1_1K^2_1K^1_2K^2_2$.
The image $w(\mathcal{L}) = 0'K^1_1L^2_1K^1_2K^2_2 \vee i
0'L^1_1K^2_1L^1_2L^2_2$ swaps the elements that appear in the nihil
part of $w$ and keeps unaltered those that appear in its certainty
part. Swaps summarize the dynamics of a production, independently of
its left hand side. Notice that, because swaps are self-adjoint, it is
enough to keep track of the certainty or nihil parts. So one
production is fully specified by, for example, its left hand side and
the nihil part of its associated swap.\footnote{Given a swap and a
  complex term $\mathcal{L}$, it is straightforward to calculate the
  production having $\mathcal{L}$ as left hand side and whose actions
  agree with those of the swap.}

We can reinterpret actions specified by productions in Matrix Graph
Grammars under MCL: instead of adding and deleting elements, they
interchange elements between the certainty and nihil parts.

The geometrical intepretation of the actions of swap $w_1 = P(p) =
0'0\wideparen{1}_2 \vee i 0'1_2$ (first element is swapped and the
rest remain unaltered)\footnote{The notation $\wideparen{1}$ stands
  for ``as many ones as nodes''. In this contribution we deal with
  finite graphs.} is a reflection with respect to the diagonal $y = x$
in the Sierpinski gasket $\mathcal{S}$. See the right of Fig.
\ref{fig:imageP}. We are considering those Boolean complexes that are
complementary in the places where the swap interchanges
elements.\footnote{For example, if $w_1$ acts on $z_1 = 0'0110 \vee i
  0'0001$ then $w_1(z_1) = z_1$. If the first element in $z_1$ is the
  same in both graphs (in fact zero) then $z_1$ is a fixed element of
  $w_1$.} Swap $w_2 = 0'10\wideparen{1}_2 \vee i 0'01_2$ divides
$\mathcal{S}$ into two regions, $S_1$ and $S_2$, along the line $y=x$.
It acts again as a reflection, but independently in $S_1$ and $S_2$.
The reflection is defined with respect to the line $y = x -
\frac{1}{2}$ in $S_1$ and with respect to $y = x + \frac{1}{2}$ in
$S_2$. The swap $w_{12} = 0'00\wideparen{1}_2 \vee i 0'01$ is the
composition of $w_1$ and $w_2$. The order does not matter.
Geometrically, $w_{12}$ corresponds to a reflection with respect to
$y=x$ and another reflection with respect to $y = x \pm \frac{1}{2}$
(only one of them applies).

\begin{figure}[htbp]
  \centering
  \includegraphics[scale = 0.3]{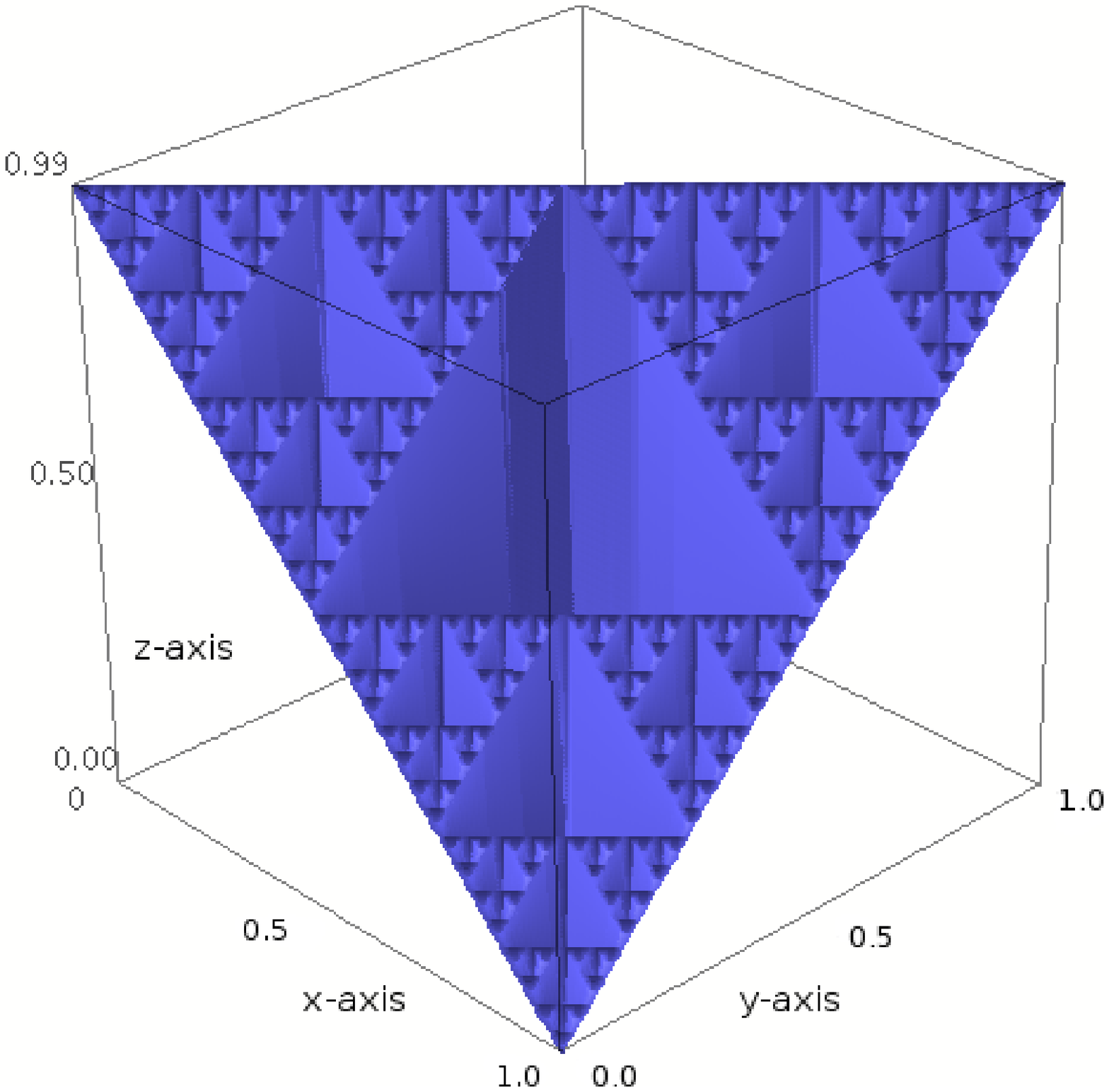}
  \includegraphics[scale = 0.3]{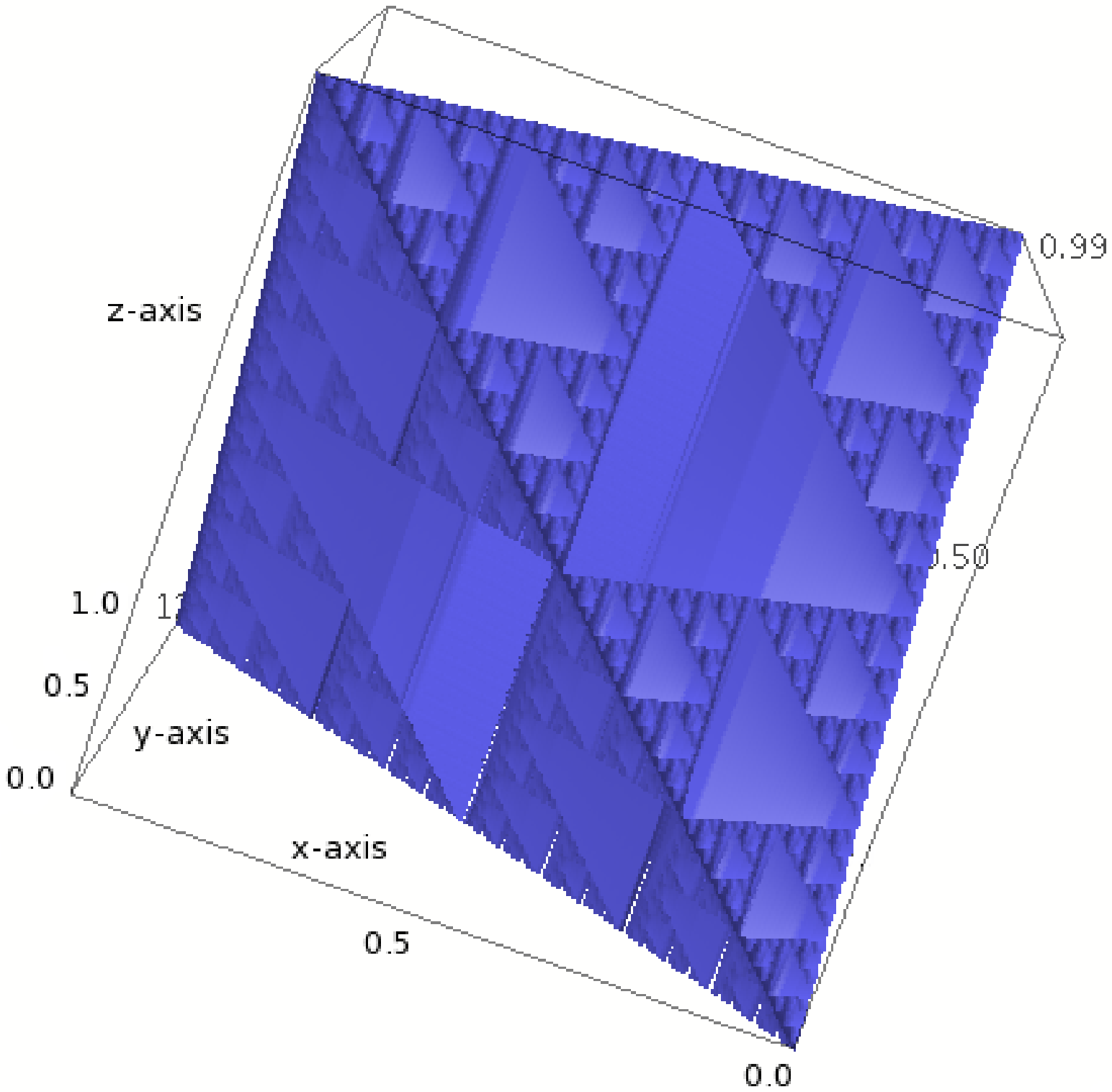}
  \caption{Image of Simple Digraphs via Swaps}
  \label{fig:nXorSwap}
\end{figure}

Let's consider those digraphs $L$ whose certainty and nihil parts are
complementary. Figure \ref{fig:nXorSwap} has $L$ on the $x$-axis and
the certainty part of the swaps $w$ in the $y$-axis. The $z$-axis
represents its image $w(L) = \overline{w \oplus L}$. Both figures are
the same surface, but the second has been rotated $\pi/4$ radians
clockwise around the z-axis.  Notice that the more elements of $L$
that we fix as zero (instead of as the complement of its certainty
part) the closer $\overline{w \oplus L}$ gets to $w \wedge L$.




\section{Coherence and Initial Digraph}
\label{sec:coherenceAndInitialDigraph}

So far we have extended MGGs by defining the transformations in
$\mathfrak{G}$ and $\mathfrak{H}$. The theory will be more interesting
if we are able to develop the necessary concepts to deal with
sequences of applications rather than productions alone. Among the two
most basic notions are coherence and the initial digraph. We shall
reformulate and extend the concepts introduced in \cite{MGG_Book}.

Recall that coherence of the sequence $s = p_n; \ldots; p_1$
guarantees that the actions of one production $p_i$ do not prevent the
actions of those sequentially behind it: $p_{i+1}, \ldots, p_{n}$ (the
first production to be applied in $s$ is $p_1$ and the last one is
$p_n$; the order is as in composition, from right to left).

\begin{theorem}[Coherence]\label{th:coherence}
  The sequence of productions $s = p_n; \ldots; p_1$ is coherent if
  the Boolean complex $C \equiv C^+ \!\vee i C^- = 0$, where
  \begin{equation}
    \label{eq:12}
    C^+ = \bigvee_{j=1}^n \left( R_j \bigtriangledown_{j+1}^n \left(
        \overline{e}_x r_y \right) \vee L_j \bigtriangleup_1^{j-1}
      \left( e_y \, \overline{r}_x \right) \right)
  \end{equation}
  and
  \begin{equation}
    \label{eq:19}
    C^- = \bigvee_{j=1}^n \left( Q_j \bigtriangledown_{j+1}^n \left(
        e_y \, \overline{r}_x \right) \vee K_j \bigtriangleup_1^{j-1}
      \left( r_y \, \overline{e}_x \right)
    \right).
  \end{equation}
\end{theorem}

\noindent\emph{Proof}\\
$\square$$C^+ \!\vee iC^- = 0 \Leftrightarrow C^+ = C^- = 0$. The
certainty part $C^+$ is addressed in \cite{MGG_Book} and $C^- = 0$ can
be proved similarly. The reader is invited to consult the proof of Th.
4.3.5 in \cite{MGG_Book} plus Lemma 4.3.3 and the explanations that
follow Def. 4.3.2 in the same reference. Next, a sequence of two
productions $s = p_2;p_1$ is considered to show the way to proceed for
$C^-$.

In order to decide whether the application of $p_1$ does not exclude
$p_2$ (regarding elements that appear in the nihil parts) the
following conditions must be demanded:
\begin{enumerate}
\item No common element is deleted by both productions:
  \begin{equation}
    \label{eq:20}
    e_1 e_2 = 0.
  \end{equation}
\item Production $p_2$ does not delete any element that the production
  $p_1$ demands not to be present and that besides is not added by
  $p_1$:
  \begin{equation}
    \label{eq:21}
    e_2 K_1 \overline{r}_1 = 0.
  \end{equation}
\item The first production does not add any element that is demanded
  not to exist by the second production:
  \begin{equation}
    \label{eq:23}
    r_1 K_2 = 0.
  \end{equation}
\end{enumerate}

Altogether we can write $e_1 e_2 \vee \overline{r}_1 e_2 K_1 \vee r_1
K_2 = e_2 (e_1 \vee \overline{r}_1 K_1) \vee r_1 K_2 = e_2 Q_1 \vee
r_1 K_2 = 0$, which is equivalent to
\begin{equation}
  \label{eq:44}
  e_2 \overline{r}_2 Q_1 \vee \overline{e}_1 r_1 K_2 = 0
\end{equation}
due to basic properties of MGG productions (see Prop. 4.1.4 in
\cite{MGG_Book}). For a sequence that consists of three productions,
$s = p_3;p_2;p_1$, the procedure is to apply the same reasoning to
subsequences $p_2;p_1$ (restrictions on $p_2$ actions due to $p_1$)
and $p_3;p_2$ (restrictions on $p_3$ actions due to $p_1$) and
\textbf{or} them. Finally, we have to deduce what has to be imposed on
$p_3$ actions due to $p_1$, but this time taking into account that
$p_2$ is applied in between. Altogether:
\begin{equation}
  \label{eq:36}
  Q_1 \left( e_2 \vee \overline{r}_2 e_3 \right) \vee Q_2 e_3 \vee
  K_2 r_1 \vee K_3 \left( r_1 \overline{e}_2 \vee r_2 \right).
\end{equation}

We may proceed similarly for four productions. Equation \eqref{eq:19}
can be deduced applying induction on the number of productions.

To see that eq. \eqref{eq:44} implies coherence we only need to
enumerate all possible actions on the nihil parts. It might be easier
if we think in terms of the negation of a potential host graph to
which both productions would be applied $\left( \overline{G} \right)$
and check that any problematic situation is ruled out. See table
\ref{tab:actionsForTwoProductions} where $D$ is deletion of one
element from $\overline{G}$ (i.e., the element is added to $G$), $A$
is addition to $G$ and $P$ is preservation.\footnote{Preservation
  means that the element is demanded to be in $\overline{G}$ because
  it is demanded not to exist by the production (it appears in $K_1$)
  and it remains as non-existent after the application of the
  production (it appears also in $Q_1$).} For example, action
$A_2;A_1$ tells that in first place $p_1$ adds one element
$\varepsilon$ to $\overline{G}$.  To do so this element has to be in
$e_1$ (or be incident to a node that is going to be deleted). After
that, $p_2$ adds the same element, deriving a conflict between the
rules.

\begin{table}[htbp]
  \centering
  \begin{tabular}{||c|c||c|c||c|c||}
    \hline
    \phantom{HH} $D_2;D_1$ \phantom{HH} & \phantom{H} \eqref{eq:23}
    \phantom{H} & \phantom{HH} $D_2;P_1$ \phantom{HH} & \phantom{H}
    $\surd$ \phantom{H} & \phantom{HH} $D_2;A_1$ \phantom{HH} &
    \phantom{H} $\surd$ \phantom{H} \\
    \hline
    $P_2;D_1$ & \eqref{eq:23} & $P_2;P_1$ & $\surd$ & $P_2;A_1$ &
    $\surd$ \\ 
    \hline
    $A_2;D_1$ & $\surd$ & $A_2;P_1$ & \eqref{eq:21} & $A_2;A_1$ &
    \eqref{eq:20} \\
    \hline
  \end{tabular}
  \caption{Possible Actions for Two Productions}
  \label{tab:actionsForTwoProductions}
\end{table}

This proves $C^- = 0$ for the case $n = 2$. When the sequence has
three productions, $s = p_3;p_2;p_1$, there are 27 possible
combinations of actions. However, some of them are considered in the
subsequences $p_2;p_1$ and $p_3;p_2$. Table
\ref{tab:actionsForThreeProductions} summarizes them.

\begin{table}[htbp]
  \centering
  \begin{tabular}{||c|c||c|c||c|c||}
    \hline
    \phantom{H} $D_3;D_2;D_1$ \phantom{H} & \phantom{H}
    \eqref{eq:23} \phantom{H} & \phantom{H} $D_3;D_2;P_1$
    \phantom{H} & \phantom{H} \eqref{eq:23} \phantom{H} &
    \phantom{H} $D_3;D_2;A_1$ \phantom{H} & \phantom{H}
    \eqref{eq:23} \phantom{H} \\
    \hline
    \phantom{H} $P_3;D_2;D_1$ \phantom{H} & \phantom{H}
    \eqref{eq:23} \phantom{H} & \phantom{H} $P_3;D_2;P_1$
    \phantom{H} & \phantom{H} \eqref{eq:23} \phantom{H} &
    \phantom{H} $P_3;D_2;A_1$ \phantom{H} & \phantom{H}
    \eqref{eq:23} \phantom{H} \\
    \hline
    \phantom{H} $A_3;D_2;D_1$ \phantom{H} & \phantom{H}
    \eqref{eq:23} \phantom{H} & \phantom{H} $A_3;D_2;P_1$
    \phantom{H} & \phantom{H} $\surd$ \phantom{H} & \phantom{H}
    $A_3;D_2;A_1$ \phantom{H} & \phantom{H} $\surd$ \phantom{H} \\
    \hline
    \phantom{H} $D_3;P_2;D_1$ \phantom{H} & \phantom{H}
    \eqref{eq:23} \phantom{H} & \phantom{H} $D_3;P_2;P_1$
    \phantom{H} & \phantom{H} $\surd$ \phantom{H} & \phantom{H}
    $D_3;P_2;A_1$ \phantom{H} & \phantom{H} $\surd$ \phantom{H} \\
    \hline
    \phantom{H} $P_3;P_2;D_1$ \phantom{H} & \phantom{H}
    \eqref{eq:23} \phantom{H} & \phantom{H} $P_3;P_2;P_1$
    \phantom{H} & \phantom{H} $\surd$ \phantom{H} & \phantom{H}
    $P_3;P_2;A_1$ \phantom{H} & \phantom{H} $\surd$ \phantom{H} \\
    \hline
    \phantom{H} $A_3;P_2;D_1$ \phantom{H} & \phantom{H}
    \eqref{eq:23}/\eqref{eq:21} \phantom{H} & \phantom{H} $A_3;P_2;P_1$
    \phantom{H} & \phantom{H} \eqref{eq:21} \phantom{H} & \phantom{H}
    $A_3;P_2;A_1$ \phantom{H} & \phantom{H} \eqref{eq:21} \phantom{H} \\
    \hline
    \phantom{H} $D_3;A_2;D_1$ \phantom{H} & \phantom{H}
    $\surd$ \phantom{H} & \phantom{H} $D_3;A_2;P_1$
    \phantom{H} & \phantom{H} \eqref{eq:21} \phantom{H} & \phantom{H}
    $D_3;A_2;A_1$ \phantom{H} & \phantom{H} \eqref{eq:20} \phantom{H} \\
    \hline
    \phantom{H} $P_3;A_2;D_1$ \phantom{H} & \phantom{H}
    $\surd$ \phantom{H} & \phantom{H} $P_3;A_2;P_1$
    \phantom{H} & \phantom{H} \eqref{eq:21} \phantom{H} & \phantom{H}
    $P_3;A_2;A_1$ \phantom{H} & \phantom{H} \eqref{eq:20} \phantom{H} \\
    \hline
    \phantom{H} $A_3;A_2;D_1$ \phantom{H} & \phantom{H}
    \eqref{eq:20} \phantom{H} & \phantom{H} $A_3;A_2;P_1$
    \phantom{H} & \phantom{H} \eqref{eq:20} \phantom{H} & \phantom{H}
    $A_3;A_2;A_1$ \phantom{H} & \phantom{H} \eqref{eq:20} \phantom{H} \\
    \hline
  \end{tabular}
  \caption{Possible Actions for Three Productions}
  \label{tab:actionsForThreeProductions}
\end{table}

There are four forbidden actions:\footnote{Those actions appearing in
  table \ref{tab:actionsForTwoProductions} updated for $p_3$.}
$D_3;D_1$, $A_3;P_1$, $P_3;D_1$ and $A_3;A_1$. Let's consider the
first one, which corresponds to $r_1 r_3$ (the first production adds
the element -- it is erased from $\overline{G}$ -- and the same for
$p_3$). In Table \ref{tab:actionsForThreeProductions} we see that
related conditions appear in positions $(1,1)$, $(4,1)$ and $(7,1)$.
The first two are ruled out by conflicts detected in $p_2;p_1$ and
$p_3;p_2$, respectively. We are left with the third case which is in
fact allowed. The condition $r_3 r_1$ taking into account the presence
of $p_2$ in the middle in eq. \eqref{eq:36} is contained in $K_3 r_1
\overline{e}_2$, which includes $r_1 \overline{e}_2 r_3$. This must be
zero, i.e. it is not possible for $p_1$ and $p_3$ to remove from
$\overline{G}$ one element if it is not added to $\overline{G}$ by
$p_2$. The other three forbidden actions can be checked similarly.

The proof can be finished by induction on the number of productions.
The induction hypothesis leaves again four cases: $D_n;D_1$,
$A_n;P_1$, $P_n;D_1$ and $A_n;A_1$. The corresponding table changes
but it is not difficult to fill in the details.$\blacksquare$

There are some duplicated conditions, so it could be possible to
``optimize'' $C$. The form considered in Th. \ref{th:coherence} is
preferred because we may use $\bigtriangleup$ and $\bigtriangledown$
to synthesize the expressions. Notice that eq. \eqref{eq:20} is
already in $C$ through eq. \eqref{eq:12}, which demands $e_1 L_2 = 0$
(as $e_2 \subset L_2$ we have that $e_1 L_2 = 0 \Rightarrow e_1 e_2 =
0$). Condition \eqref{eq:21} is $e_2 K_1 \overline{r}_1 = e_2
\overline{r}_1 r_1 \vee e_2 \overline{r}_1 \overline{e}_1
\overline{D}_1 = e_2 \overline{e}_1 \overline{D}_1$, where we have
used that $K_1 = p \left( \overline{D}_1 \right)$. Note that those
$\overline{e}_1 \overline{D}_1 \neq 0$ are the dangling edges not
deleted by $p_1$. Finally, equality \eqref{eq:23} is $r_1 K_2 = r_1
p_2 \left( \overline{D}_2 \right) = r_1 \left( r_2 \vee \overline{e}_2
  \overline{D}_2 \right) = r_1 r_2 \vee r_1 \overline{e}_2
\overline{D}_2$. The first term $(r_1 r_2)$ is already included in $C$
and the second term is again related to dangling edges. Potential
dangling edges appear in coherence and this may seem to indicate a
possible link between coherence and
compatibility.\footnote{Compatibility for sequences is characterized
  in Sec.  \ref{sec:compatibilityAndCongruence}. Coherence takes into
  account dangling edges, but only those that appear in the
  ``actions'' of the productions (in matrices $e$ and $r$).}

\begin{figure}[htbp]
  \centering
  \includegraphics[scale = 0.5]{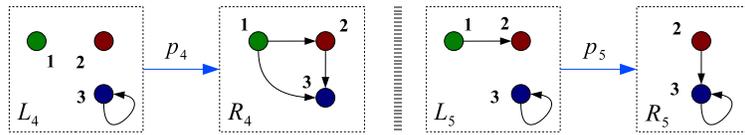}
  \caption{Example of Coherence}
  \label{fig:ex5}
\end{figure}

\textbf{Example}. Let's consider the sequence $s = p_5;p_4$. Recall
that the order of application is from right to left so $p_4$ is
applied first and $p_5$ right afterwards. Let $p_4$ and $p_5$ be those
productions depicted in Fig. \ref{fig:ex5}. Once simplified, its
coherence term is
\begin{eqnarray}
  \label{eq:38}
  C(s) & = & C^+(s) \vee i C^-(s) = \left( R_4 r_5 \vee L_5 e_4 \right) \vee i
  \left( Q_4 e_5 \vee K_5 r_4 \right) = \nonumber \\
  & = & \left( \left[ \begin{array}{ccc}
        0 & 1 & 1 \\ 
        0 & 0 & 1 \\
        0 & 0 & 0
      \end{array} \right] \left[ \begin{array}{ccc}
        0 & 0 & 0 \\ 
        0 & 0 & 1 \\
        0 & 0 & 0
      \end{array} \right] \vee \left[ \begin{array}{ccc}
        0 & 1 & 0 \\ 
        0 & 0 & 0 \\
        0 & 0 & 1
      \end{array} \right] \left[ \begin{array}{ccc}
        0 & 0 & 0 \\ 
        0 & 0 & 0 \\
        0 & 0 & 1
      \end{array} \right] \right) \vee i \left( \left[
      \begin{array}{ccc} 
        0 & 0 & 0 \\ 
        0 & 0 & 0 \\
        0 & 0 & 1
      \end{array} \right] \left[ \begin{array}{ccc}
        0 & 1 & 0 \\ 
        0 & 0 & 0 \\
        0 & 0 & 0
      \end{array} \right] \vee \right. \nonumber \\
  & & \left. \vee \left[ \begin{array}{ccc}
        1 & 0 & 1 \\ 
        1 & 0 & 1 \\
        1 & 0 & 0
      \end{array} \right] \left[ \begin{array}{ccc}
        0 & 1 & 1 \\ 
        0 & 0 & 1 \\
        0 & 0 & 0
      \end{array} \right] \right) = \left[ \begin{array}{ccc}
      0 & 0 & 0 \\ 
      0 & 0 & 1 \\
      0 & 0 & 1
    \end{array} \right] \vee i \left[ \begin{array}{ccc}
      0 & 0 & 1 \\ 
      0 & 0 & 1 \\
      0 & 0 & 0
    \end{array} \right]. \nonumber
\end{eqnarray}

Coherence problems appear in this example for several reasons. Edge
$(2,3)$ is added twice while self-loop $(3,3)$ is first deleted in
$p_4$ and then used in $p_5$. Edge $(1,3)$ becomes dangling because
production $p_5$ deletes node $1$. Edge $(2,3)$ appears in $C^-(s)$
for the same reason that makes it appear in $C^+(s)$.$\blacksquare$

The minimal initial digraph $M(s)$ for a completed sequence $s = p_n;
\ldots; p_1$ was introduced in \cite{MGG_Book} as a simple digraph
that permits all operations of $s$ and that does not contain a proper
subgraph with the same property. The negative initial digraph has a
similar definition but for the nihil part. Theorem \ref{prop:seqComp}
encodes as a complex term the minimal and negative initial digraphs,
renaming it to \emph{initial digraph}.

Now we are interested in what elements will be forbidden and which
ones will be available once every production is
applied.\footnote{Recall that whenever the tensor (Kronecker) product
  is used, we refer to the vector of nodes so the $V$ superscript is
  omitted. For example $R \otimes R' \equiv R^V \otimes \left( R^V
  \right)^t$. The ${}^t$ stands for transposition.} Matrix
$\overline{D} = \overline{\overline{e} \otimes \overline{e}^t}$
specifies what edges can not be present because at least one of their
incident nodes have been deleted. Let's introduce the dual concept:
\begin{equation}
  \label{eq:42}
  T = \left( \overline{ \overline{r} \otimes \overline{r}^t} \right)
  \wedge \left( \overline{e} \otimes \overline{e}^t \right).
\end{equation}

$T$ are the newly available edges after the application of the
production because of the addition of nodes.\footnote{This is why $T$
  does not appear in the calculation of the coherence of a sequence:
  coherence takes care of real actions $(e,r)$ and not of potential
  elements that may or may not be available $\left( \overline{D}, T
  \right)$.} The first term, $\overline{ \overline{r} \otimes
  \overline{r}^t}$, has a one in all edges incident to a vertex that
is added by the production. We have to remove those edges that are
incident to some node deleted by the production, which is what
$\overline{e} \otimes \overline{e}^t$ does.

\begin{figure}[htbp]
  \centering
  \includegraphics[scale = 0.58]{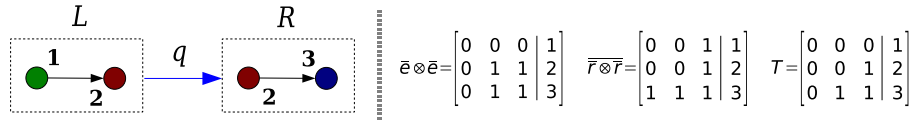}
  \caption{Available and Unavailable Edges After the Application of a
    Production}
  \label{fig:matrixT}
\end{figure}

\noindent \textbf{Example}.$\square$Figure \ref{fig:matrixT} depicts
to the left a production $q$ that deletes node $1$ and adds node $3$.
Its nihil term and its image are
\begin{equation}
  \label{eq:17}
  K = q \left( \overline{D} \right) = r \vee \overline{e} \overline{D}
  = \left[
    \begin{array}{ccc}
      1 & 0 & 1 \\ 
      1 & 0 & 1 \\
      1 & 0 & 0
    \end{array}
  \right] \qquad Q = q^{-1}(K) = e \vee \overline{r}K = \left[
    \begin{array}{ccc}
      1 & 1 & 1 \\ 
      1 & 0 & 0 \\
      1 & 0 & 0
    \end{array}
  \right] \nonumber
\end{equation}

To the right of Fig. \ref{fig:matrixT} matrix $T$ is included. It
specifies those elements that are not forbidden once production $q$
has been applied.$\blacksquare$

Matrices $\overline{D}$ and $T$ do not tell actions of the production
to be performed in the complement of the host graph, $\overline{G}$.
Actions of productions are specified exclusively by matrices $e$ and
$r$.

\begin{theorem}[Initial Digraph]\label{th:ID}
  The \emph{initial digraph} $M(s)$ for the completed coherent
  sequence of productions $s = p_n; \ldots; p_1$ is given by
  \begin{equation}
    \label{eq:24}
    M(s) = \bigtriangledown_1^n \left( \overline{r}_x L_y \vee i \, 
      \overline{e}_x \overline{T}_x K_y \right).
  \end{equation}
\end{theorem}

\noindent \emph{Proof (sketch)} \\
$\square$The proof proceeds along the lines of that for Th. 4.4.2 in
\cite{MGG_Book}, which in essence starts with a big enough graph and
removes as many elements as possible. However, for edges, besides the
actions of the productions on edges we need to keep track of the
actions of the productions on nodes because some potential dangling
edges may become available (if their incident nodes are added by some
grammar rule).

Notice that Th. 4.4.2 in \cite{MGG_Book} proves that the certainty
part of the initial digraph is the one that appears in eq.
\eqref{eq:24}. For the nihil term $\bigtriangledown_1^n \overline{e}_x
\overline{T}_x K_y$ it is easier to think in what must be or must not
be found in $\overline{G}$.

We proceed by induction on the number of productions. For the time
being, for simplicity, we omit the effect of adding nodes which may
turn potential dangling edges into available ones. In a sequence with
a single production it should be obvious that $K_1$ (and only $K_1$)
needs to be demanded.

For a sequence of two productions $s_2 = p_2;p_1$, $K_1$ is again
necessary. It is clear that $K_1 \vee K_2$ with $K_1^V K_2^V = 0$ --
i.e. all nodes and hence edges unrelated -- would be enough, but it
may include more elements than strictly needed. Among them, those
already deleted by $p_1$ (once they are deleted they belong to
$\overline{G}$) and those that already appear in $K_1$ and that are
not added by $p_1$ -- $\overline{r_1} K_1$ --. If these elements of
$K_2$ are not going to be considered, we need to \textbf{and} their
negation: $\overline{e}_1 \overline{\left( \overline{r}_1 K_1 \right)}
K_2$.  Altogether we get $K_1 \vee \overline{e}_1 \overline{ \left(
    \overline{r}_1 K_1 \right)} K_2$. Some simple manipulations prove
that:
\begin{eqnarray}
  \label{eq:32}
  K_1 \vee K_2 \overline{e}_1 \overline{\left( \overline{r}_1 K_1
    \right) } & = & K_1 \vee K_2 \overline{e}_1 \left( r_1 \vee
    \overline{K}_1 \right) = \nonumber \\
  & = & K_1 \vee K_2 \overline{ \left( e_1 \vee \overline{r}_1 K_1
    \right)} = K_2 \overline{Q}_1.
\end{eqnarray}

Minimality is inferred by construction. If any other element was
removed then either $p^{-1}_1$ or $p^{-1}_2$ could not be applied (and
still consider dangling edges). It is not difficult to check that the
sequence $p^{-1}_2; p^{-1}_1$ can be applied to $K_1 \vee K_2
\overline{Q_1}$.  The expressions for sequences of three, four, $n$
poductions are:
\begin{eqnarray}
  N_3 & = & N_2 \vee K_3 \overline{ \overline{r}_2 Q_1} \,
  \overline{Q}_2 \\
  N_4 & = & N_3 \vee K_4 \overline{\overline{r}_3 \overline{r}_2 Q_1}
  \, \overline{\overline{r}_2 Q_2} \, \overline{Q}_3 \\
  \ldots & & \nonumber \\
  \label{eq:34}
  N_n & = & K_1 \vee \overline{r}_1 \bigtriangledown_1^{n-1} \left(
    \overline{Q}_x K_{y+1} \right) \vee \bigvee_{j=2}^n \left[ K_j
    \bigtriangleup_1^{j-1} \left( \overline{Q}_x r_y \right) \right]
  \\
  \label{eq:35}
  N_n & = & K_1 \vee \overline{r_1} \bigtriangledown_1^{n-1} \left(
    \overline{e}_x K_{y+1} \right) \vee \bigvee_{j=2}^n \left[ K_j
    \bigtriangleup_1^{j-1} \left( \overline{e}_x r_y \right)
  \right].
\end{eqnarray}

There are two tricky steps. The first one is how to derive $N_n$ in
eq. \eqref{eq:34} and the second is how to obtain its equivalent
expression \eqref{eq:35}. The reader is referred again to the
aforementioned proof in \cite{MGG_Book} where detailed explanations
are given for a similar case.

Once we get here it is easy to obtain $\bigtriangledown_1^n \left(
  \overline{e}_x K_y \right)$. First, note that the sequence is
coherent so the third term in eq. \eqref{eq:35} is zero. Second, as
$K_1 = K_1 \vee r_1$, the $\overline{r}_1$ can be simplified because
$a \vee \overline{a}b = a \vee b$ in propositional logic.

Finally, the same reasoning applies for those nodes that are added. So
we do not only need to remove elements erased by previous productions
but also edges that are not incident to any non-existent edge,
$\bigtriangledown_1^n \left( \overline{e}_x K_y \right) \mapsto
\bigtriangledown_1^n \left( \overline{e}_x \overline{T}_x K_y
\right)$.$\blacksquare$

\begin{figure}[htbp]
  \centering
  \includegraphics[scale = 0.57]{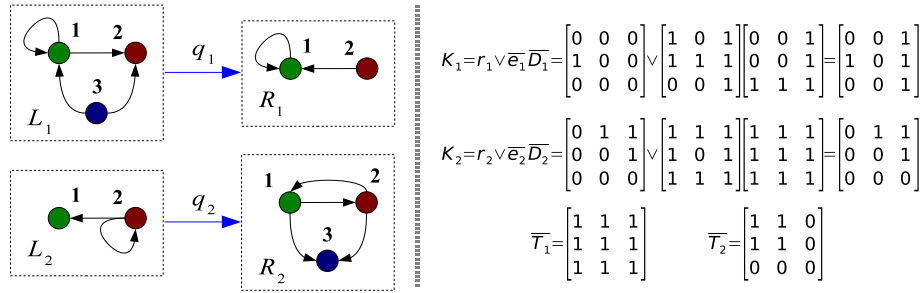}
  \caption{sequence of Two Productions}
  \label{fig:ex2}
\end{figure}

\noindent \textbf{Example}.$\square$Figure \ref{fig:ex2} includes two
productions with their nihilation matrices $K_1$ and $K_2$. The
initial digraph of the sequence $s = q_2;q_1$ is
\begin{eqnarray}
  \label{eq:27}
  M(s_2) & = & \bigtriangledown_1^2 \left( \overline{r}_x L_y \! \vee
    \! i \overline{e}_x \overline{T}_x K_y \right) = \left(
    \overline{r}_1 L_1 \! \vee \! \overline{r}_1 \overline{r}_2
    L_2 \right) \! \vee \! i \left( \overline{e}_1 \overline{T}_1 K_1
    \! \vee \! \overline{e}_1 \overline{e}_2 \overline{T}_1
    \overline{T}_2 K_2 \right) = \nonumber \\
  & = & \left( L_1 \vee \overline{r}_1 L_2 \right) \! \vee \! i \left(
    \overline{T}_1 K_1 \vee \overline{e}_1 \overline{T}_1
    \overline{T}_2 K_2 \right) \! = \! \left( \! \left[
      \begin{array}{ccc}
        1 & 1 & 0 \\ 
        0 & 0 & 0 \\
        1 & 1 & 0
      \end{array} \right] \!\! \vee \!\! \left[ \begin{array}{ccc}
        1 & 1 & 1 \\ 
        0 & 1 & 1 \\
        1 & 1 & 1
      \end{array} \right] \!\! \left[ \begin{array}{ccc}
        0 & 0 & 0 \\ 
        1 & 1 & 0 \\
        0 & 0 & 0
      \end{array} \right] \! \right) \nonumber \\
  & & \vee i \left( \left[ \begin{array}{ccc}
        1 & 1 & 1 \\ 
        1 & 1 & 1 \\
        1 & 1 & 1
      \end{array} \right] \!\! \left[ \begin{array}{ccc}
        0 & 0 & 1 \\ 
        1 & 0 & 1 \\
        0 & 0 & 1
      \end{array} \right] \!\! \vee \!\! \left[ \begin{array}{ccc}
        1 & 0 & 1 \\ 
        1 & 1 & 1 \\
        0 & 0 & 1
      \end{array} \right] \!\! \left[ \begin{array}{ccc}
        1 & 1 & 1 \\ 
        1 & 1 & 1 \\
        1 & 1 & 1
      \end{array} \right] \!\! \left[ \begin{array}{ccc}
        0 & 0 & 1 \\ 
        0 & 0 & 1 \\
        1 & 1 & 1
      \end{array} \right] \!\! \left[ \begin{array}{ccc}
        0 & 1 & 1 \\ 
        0 & 0 & 1 \\
        0 & 0 & 0
      \end{array} \right] \right) =  \nonumber \\
  & = & \left[ \begin{array}{ccc}
      1 & 1 & 0 \\ 
      0 & 1 & 0 \\
      1 & 1 & 0
    \end{array} \right] \vee i \left[ \begin{array}{ccc}
      0 & 0 & 1 \\ 
      1 & 0 & 1 \\
      0 & 0 & 1
    \end{array} \right] \equiv M_C(s_2) \vee M_N(s_2).\nonumber
\end{eqnarray}

\begin{figure}[htbp]
  \centering
  \includegraphics[scale = 0.44]{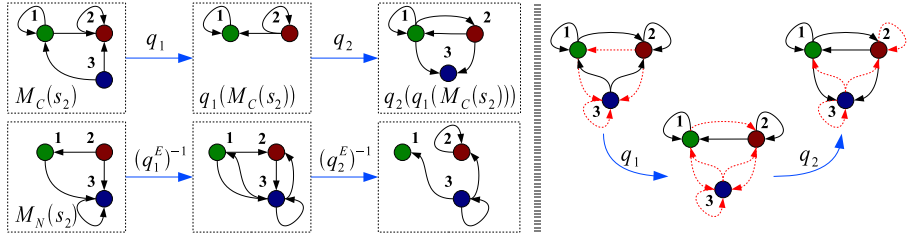}
  \caption{Initial Digraph of a Sequence of Two Productions Together
    with its Evolution}
  \label{fig:ex3}
\end{figure}

We have represented $M_C(s_2) \vee M_N(s_2)$ to the left of Fig.
\ref{fig:ex3} together with its evolution as well as the final state,
$s_2 \left( M(s_2) \right)$. To the right of the same figure there is
the same evolution but limited to edges and from the point of view of
swaps. With black solid line we have represented the edges that are
present and with red dotted line those that are absent. 
Recall that swaps interchange them.$\blacksquare$

A final remark is that $\overline{T}$ makes the number of edges in
$\overline{G}$ as small as possible. For example, in $\overline{e}_1
\overline{e}_2 \overline{T}_1 \overline{T}_2 K_2$ we are in particular
demanding $\overline{e}_1 \overline{T}_1 \overline{T}_2 r_2$ (because
$K_2 = r_2 \vee \overline{e}_2 \overline{D}_2$). If we start with a
compatible host graph, it is not necessary to ask for the absence of
edges incident to nodes that are added by a production (we called them
\emph{potentially available} above). Notice that these edges could not
be in the host graph as they would be dangling edges or we would be
adding an already existent node).

\section{Compatibility and Congruence}
\label{sec:compatibilityAndCongruence}

This section revises some more sequential results, adapting and
extending them via MCL. The notions we cope with are the image of a
sequence, compatibility\footnote{Compatibility for a single production
  has been tackled in Prop. \ref{prop:production}, Sec.
  \ref{sec:numericalRepresentation}.} and G-congruence. By the end of
the section we will very briefly touch on sequential independence,
application conditions and graph constraints.

The image of a sequence of productions $s = p_n; \ldots; p_1$ acting
on its initial digraph $M(s) = M_C(s) \vee i M_N(s)$ is given by:
\begin{equation}
  s \left( M(s) \right) = \left( \bigtriangleup_1^n \overline{e}_x r_y
    \vee \bigwedge_{j=1}^n \overline{e}_j M_C(s) \right) \vee i
  \left( \bigtriangleup_1^n \overline{r}_x e_y \vee \bigwedge_{j=1}^n
    \overline{r}_j M_N(s) \right).
  \label{eq:43}
\end{equation}

Equation \eqref{eq:43} is deduced by simply applying each production
to the initial digraph.  We would like to interpret the shape of the
image of a sequence as a production by setting $s = e(s) \vee i r(s)$,
with $e(s) = \bigvee_{i=1}^n e_i$ and $r(s)= \bigtriangleup_1^n
\overline{e}_x r_y$.\footnote{The idea behind would be the composition
  of a sequence of productions to derive a single production: $s_n =
  p_n; \ldots; p_1 \mapsto c_n = p_n \circ \ldots \circ p_1$.} We
could then put it as $\left \langle M(s), s \right \rangle$ and for
example calculate its associated swap:
\begin{equation}
  w(s) = \bigwedge_{j=1}^n \left( \overline{e}_j \overline{r}_j
  \right) \vee i \bigvee_{j=1}^n \left( e_j \vee r_j
  \right). \nonumber
\end{equation}

Unfortunately this is not possible because although in the certainty
part of eq. \eqref{eq:43} $r(s) = \bigtriangleup_1^n \overline{e}_x
r_y$ and $e(s) = \bigvee_{i=1}^n e_i$, in the nihil part we find
$\bigwedge_{j=1}^n \overline{r}_j \neq \overline{r(s)}$ and
$\bigtriangleup_1^n \overline{r}_x e_y \neq \overline{e(s)}$.

Compatibility asks for ``closedness'' of the space (graphs) with
respect to the specified operations. In essence, it demands the lack
of dangling edges. Definitions of compatibility for increasingly
general concepts can be found in \cite{MGG_Book}: single simple
digraph, production and sequence. According to Prop.
\ref{prop:production} productions act on edges and on vertices. They
are obviously related but this relation has not been demonstrated. It
is of importance in order to study the evolution of the nihil part of
complex terms. What one production forbids, another production may
need or even can make accessible again.

\begin{proposition}[Compatibility]\label{prop:seqComp}
  Let $s = p_n; \ldots; p_1$ be a sequence made up of compatible
  productions. If
  \begin{equation}
    \label{eq:39}
    \bigtriangledown_1^n \left( \overline{e}_x \overline{r}_x M_C(s_x)
      M_N(s_x)\right) = 0
  \end{equation}
  then $s$ is compatible, where $M_C(s_m)$ and $M_N(s_m)$ are the
  certainty and nihil parts of the initial digraphs of $s_m = p_m;
  \ldots; p_1$, $m \in \{1, \ldots, n\}$.
\end{proposition}

\noindent \emph{Proof (Sketch)}\\
$\square$Equation \eqref{eq:39} is a restatement of the definition of
compatibility for a sequence of productions. The condition appears
when the certainty and nihil parts are demanded to have no common
elements. Compatibility of each production is used to simplify terms
of the form $L_i K_i$.$\blacksquare$

Compatibility and coherence are related notions but only to some
extent. Coherence deals with actions of productions while
compatibility with potential presence or abscense of elements. This is
better understood if we think in derivations: when the left hand side
$L \vee i K$ of rule $p$ is matched in a host graph $G \vee i
\overline{G}$, all elements of $L$ must be found in $G$ and all edges
of $K$ must be found in $\overline{G}$. When $p$ is applied a new
graph $H \vee i\overline{H}$ is derived. Again, all elements of $R$
have to be found in $H$ and all edges in $Q$ will be in
$\overline{H}$, no matter if some of them are now potentially usable
(say $p$ adds some nodes and some potentially dangling edges are not
dangling edges anymore).

Now we turn to G-congruence, which studies equality of initial
digraphs for a sequence $s = p_n; \ldots; p_1$ and a permutation of
it, $s' = \sigma(s)$. All the job for advancement and delaying of
productions -- permutations $\phi$ and $\delta$,
where\footnote{Numbers in the permutation refers to the position that
  the production occupies inside the sequence, not to its subindex.}
advancement is $\phi = (1 \; 2 \;\ldots \; n\!-\!1 \; n)$ and delaying
is $\delta = (n \; n\!-\!1 \;\ldots \; 2 \; 1)$, i.e $\phi(s) =
p_{n-1}; p_{n-2} \ldots ; p_1; p_n$ and $\delta(s) = p_1; p_n; \ldots
p_2$ -- is done in \cite{MGG_Book}, Sec. 6.1, so we state the result
without proof.

\begin{theorem}[G-congruence]\label{th:gCongruence}
  With notation as above, sequences $s$ and $\phi(s)$ are G-congruents
  if $F \equiv F^+ \vee iF^- = 0$, where
  \begin{equation}
    \label{eq:28}
    F^+ = L_n \nabla_1^{n-1} \overline{e}_x K_y \left(r_y \vee e_n
    \right) \quad \mathrm{and} \quad F^- = K_n \nabla_1^{n-1}
    \overline{r}_x L_y \left( e_y \vee r_n \right).
  \end{equation}
  Also, $s$ and $\delta(s)$ are G-congruents if $D \equiv D^+ \vee i
  D^- = 0$, with
  \begin{equation}
    \label{eq:29}
    D^+ = L_1 \nabla_2^{n} \overline{e}_x K_y \left(r_y \vee e_1
    \right) \quad \mathrm{and} \quad D^- =  K_1 \nabla_2^{n}
    \overline{r}_x L_y \left( e_y \vee r_1 \right).
  \end{equation}
\end{theorem}

\noindent \emph{Proof}\\
$\square$$\blacksquare$

An easy remark is that the complex term $C^+ \vee i C^-$ in Th.
\ref{th:coherence} provides more information than just settling
coherence as it measures non-coherence: \emph{Problematic} elements
(i.e. those that prevent coherence) would appear as ones and the rest
as zeros. The same holds for $F^+\!\vee\!i F^-$ and $D^+\!\vee\!i D^-$
in Th. \ref{th:gCongruence} for congruence and eq. \eqref{eq:39} in
Prop. \ref{prop:seqComp} for compatibility.

There are some relevant topics that we have not mentioned such as
sequential independence, application conditions and graph constraints.
We briefly discuss how they could be handled with MCL.

With respect to the image of a sequence and sequential independence,
recall that MCL naturally uses swaps rather than productions. This
abstraction has its effects on the interpretation of operations. On
the positive side, among many other things, swaps are a nice
redefinition and generalization of productions that take into account
the certainty and nihil parts; on the negative side, our intuition
needs to be adjusted. For example, consider a production $p$ that only
deletes edge $(1,2)$ and does nothing else. Suppose that it is applied
twice to the graph $G$ that consists of nodes $1$, $2$ and edge
$(1,2)$. In this case $p;p(G) = G$ which is algebraically correct.
However, it does not encode ``delete edge $(1,2)$ twice''.  Of course,
the point here is that of completion: we would rather have considered
its application to $G'$, made up of nodes $1$, $1'$ and $2$ and edges
$(1,2)$ and $(1',2)$. A similar reasoning shows that sequential
independence is ``granted'' if we rely only on algebraic operations
and do not pay attention to completion:
\begin{equation}
  \label{eq:37}
  p_2;p_1(\mathcal{L}) = \left \langle \left \langle \mathcal{L},
      P(p_1) \right \rangle, P(p_2) \right \rangle = \mathcal{L}
  P(p_1) P(p_2) = \mathcal{L} P(p_2) P(p_1) =
  p_1;p_2(\mathcal{L}).\nonumber 
\end{equation}
Previous comments highlight some of the reasons why coherence,
compatibility, initial digraph and G-congruence are so valuable,
justifying their inclusion and also linking the present and previous
sections to Sec. \ref{sec:productionEncoding}.

Regarding application conditions and graph constraints, they are not
difficulty related to what has been presented so far. Recall from Sec.
\ref{sec:productionEncoding} that swaps transform elements in the same
diagonal of the Sierpinski gasket. If they are allowed to be applied
to $g \in \mathfrak{G}$ instead of the restricted case that we have
studied $(\mathfrak{H})$, we may impose limits on what elements can
not be added nor deleted by sequences of productions (swaps). This is
because if one edge is in the certainty part and in the nihil part, it
can not be deleted by any swap. On the contrary, if one edge does not
appear neither in the certainty nor in the nihil parts, it is not
possible for a swap to add it.

If we call any of these situations a \emph{swap restriction}, it can
be guaranteed that a sequence will not add nor delete (or both) some
element, despite the actual definition of the productions that make up
the sequence or the grammar. Again, geometrically, we are choosing the
diagonal inside the Sierpinski gasket in which all operations will
take place.

\section{Conclusions and Future Work}
\label{sec:conclusions}

In this paper we have introduced \emph{Monotone Complex Logic} (MCL)
which, in our opinion, is an interesting topic in itself. With respect
to Matrix Graph Grammars (MGGs), MCL allows the encoding of simple
digraphs and grammar rules using complex terms. We believe it is a
natural representation in the MGGs context, as productions act on
pairs of graphs $(L, K) \overset{p} \rightarrow (p(L), p^{-1}(K))$.
Relevant algebraic structures for their study have been introduced
(PMCA, PMMA, $\mathfrak{G}$, $\mathfrak{H}$). Swaps allow studying and
classifying productions according to their dynamic behaviour, defining
a surjective morphism into the self-adjoint graphs in $\mathfrak{H}$.

The rational enconding of Boolean complexes gives an embedding of MGGs
into a subset of the complex numbers: the Sierpinski gasket. Using
such representation we have been able to introduce standard geometric
and analytic concepts such as a scalar product, a norm and a notion of
distance in MGGs. This generalizes the theory developed
in~\cite{MGG_Book} and opens the door to the study of dynamics of
infinite graphs with a countable number of nodes (this topic is left
for future research). Finally, some of the most relevant concepts of
MGGs have been expressed and reinterpreted using MCL: coherence,
initial digraph, image of a sequence, compatibility and G-congruence.

Our main interest is complexity theory so we have to introduce a
measure of complexity.  The natural proposal seems to be the geodesic
distance, which measures the cost of reaching one element from another
one through elements of the MGG (rules of the grammar). However, the
\emph{natural} distance here is not the Euclidean one, but the
\textbf{xor} metric restricted by available operations.

One foreseen advantage of the results in this paper is that there is a
lot of interest and current research activity on the Sierpinski
gasket~\cite{geometry,Kigami,Weisstein}. In the mid-long term we plan
to continue our work towards computational complexity theory through
MGGs. It is our opinion that one of the main ``problems'' of current
approaches to complexity theory is that there are very few links to
other branches of mathematics (there are some exceptions though, such
as~\cite{Mulmuley}). As MGGs are a compact and path connected fractal,
it seems promising to introduce harmonic and functional analysis and
noncommutative geometry techniques, apart from those already available
in MGGs.  Two main research directions will be explored in the future:
measurable Riemannian geometry as in \cite{Kigami} and noncommutative
geometry as in \cite{geometry}.

Notice that it is not difficult to interpret MGGs as a model of
computation (we are preparing a paper on this topic). Also, it might
be of interest to encode properties of graphs (such as coloring) using
graph grammars, translating \emph{static} properties into equivalent
\emph{dynamic} properties of associated sequences.

Another point of interest might be the introduction of stochastic
analysis. This is closely related to MGGs as a model of computation
and the way grammar rules are selected (a source of non-determinism).
Other source of non-determinism appears in case there are several
places in a host graph to which a production can be applied.

There are many more topics for further research, e.g. graph
constraints, derivations, applicability, reachability, dynamic
encoding of static properties and infinite graphs some of which we
have already commented on.

\textbf{Acknowledgements}: Pedro Pablo wants to thank the open source
community. \emph{SAGE} (\url{http://www.sagemath.org/}) has been used
for some calculations, in particular those necessary to generate Figs.
\ref{fig:mgg_CharSet} and \ref{fig:nXorSwap}.  \emph{OpenOffice
  Drawing} (\url{http://www.openoffice.org/}) has been used with Figs.
\ref{fig:ex4}, \ref{fig:exProds}, \ref{fig:imageP}, \ref{fig:matrixT},
\ref{fig:ex2} and \ref{fig:ex3}. \emph{The Gimp}
(\url{http://www.gimp.org/}) has helped with some finishing touches.
\emph{Emacs} (\url{http://www.gnu.org/software/emacs/}) is unvaluable
for typing and \emph{teTeX} for $\LaTeX$.


\end{document}

%% file: psfig_pp
\catcode`\@=11\relax
\newwrite\@unused
\def\typeout#1{{\let\protect\string\immediate\write\@unused{#1}}}
\def\figurepath{./}

\def\@nnil{\@nil}
\def\@empty{}
\def\@psdonoop#1\@@#2#3{}
\def\@psdo#1:=#2\do#3{\edef\@psdotmp{#2}\ifx\@psdotmp\@empty \else
    \expandafter\@psdoloop#2,\@nil,\@nil\@@#1{#3}\fi}
\def\@psdoloop#1,#2,#3\@@#4#5{\def#4{#1}\ifx #4\@nnil \else
       #5\def#4{#2}\ifx #4\@nnil \else#5\@ipsdoloop #3\@@#4{#5}\fi\fi}
\def\@ipsdoloop#1,#2\@@#3#4{\def#3{#1}\ifx #3\@nnil 
       \let\@nextwhile=\@psdonoop \else
      #4\relax\let\@nextwhile=\@ipsdoloop\fi\@nextwhile#2\@@#3{#4}}
\def\@tpsdo#1:=#2\do#3{\xdef\@psdotmp{#2}\ifx\@psdotmp\@empty \else
    \@tpsdoloop#2\@nil\@nil\@@#1{#3}\fi}
\def\@tpsdoloop#1#2\@@#3#4{\def#3{#1}\ifx #3\@nnil 
       \let\@nextwhile=\@psdonoop \else
      #4\relax\let\@nextwhile=\@tpsdoloop\fi\@nextwhile#2\@@#3{#4}}
\def\psdraft{
	\def\@psdraft{0}
}
\def\psfull{
	\def\@psdraft{100}
}
\psfull
\newif\if@prologfile
\newif\if@postlogfile
\newif\if@noisy
\def\pssilent{
	\@noisyfalse
}
\def\psnoisy{
	\@noisytrue
}
\psnoisy
\newif\if@bbllx
\newif\if@bblly
\newif\if@bburx
\newif\if@bbury
\newif\if@height
\newif\if@width
\newif\if@scale
\newif\if@rheight
\newif\if@rwidth
\newif\if@clip
\newif\if@verbose
\def\@p@@sclip#1{\@cliptrue}

\def\@p@@sfile#1{\def\@p@sfile{null}%
	        \openin1=#1
		\ifeof1\closein1%
		       \openin1=\figurepath#1
			\ifeof1\typeout{Error, File #1 not found}
			\else\closein1
			    \edef\@p@sfile{\figurepath#1}%
                        \fi%
		 \else\closein1%
		       \def\@p@sfile{#1}%
		 \fi}
\def\@p@@sfigure#1{\def\@p@sfile{null}%
	        \openin1=#1
		\ifeof1\closein1%
		       \openin1=\figurepath#1
			\ifeof1\typeout{Error, File #1 not found}
			\else\closein1
			    \def\@p@sfile{\figurepath#1}%
                        \fi%
		 \else\closein1%
		       \def\@p@sfile{#1}%
		 \fi}

\def\@p@@sbbllx#1{
		\@bbllxtrue
		\dimen100=#1
		\edef\@p@sbbllx{\number\dimen100}
}
\def\@p@@sbblly#1{
		\@bbllytrue
		\dimen100=#1
		\edef\@p@sbblly{\number\dimen100}
}
\def\@p@@sbburx#1{
		\@bburxtrue
		\dimen100=#1
		\edef\@p@sbburx{\number\dimen100}
}
\def\@p@@sbbury#1{
		\@bburytrue
		\dimen100=#1
		\edef\@p@sbbury{\number\dimen100}
}
\def\@p@@sscale#1{
		\@scaletrue
		\count255=#1
   		\edef\@p@sscale{\number\count255}
}
\def\@p@@sheight#1{
		\@heighttrue
		\dimen100=#1
   		\edef\@p@sheight{\number\dimen100}
}
\def\@p@@swidth#1{
		\@widthtrue
		\dimen100=#1
		\edef\@p@swidth{\number\dimen100}
}
\def\@p@@srheight#1{
		\@rheighttrue
		\dimen100=#1
		\edef\@p@srheight{\number\dimen100}
}
\def\@p@@srwidth#1{
		\@rwidthtrue
		\dimen100=#1
		\edef\@p@srwidth{\number\dimen100}
}
\def\@p@@ssilent#1{ 
		\@verbosefalse
}
\def\@p@@sprolog#1{\@prologfiletrue\def\@prologfileval{#1}}
\def\@p@@spostlog#1{\@postlogfiletrue\def\@postlogfileval{#1}}
\def\@cs@name#1{\csname #1\endcsname}
\def\@setparms#1=#2,{\@cs@name{@p@@s#1}{#2}}
\def\ps@init@parms{
		\@bbllxfalse \@bbllyfalse
		\@bburxfalse \@bburyfalse
		\@heightfalse \@widthfalse
		\@scalefalse
		\@rheightfalse \@rwidthfalse
		\def\@p@sbbllx{}\def\@p@sbblly{}
		\def\@p@sbburx{}\def\@p@sbbury{}
		\def\@p@sheight{}\def\@p@swidth{}
		\def\@p@sscale{}
		\def\@p@srheight{}\def\@p@srwidth{}
		\def\@p@sfile{}
		\def\@p@scost{10}
		\def\@sc{}
		\@prologfilefalse
		\@postlogfilefalse
		\@clipfalse
		\if@noisy
			\@verbosetrue
		\else
			\@verbosefalse
		\fi
}
\def\parse@ps@parms#1{
	 	\@psdo\@psfiga:=#1\do
		   {\expandafter\@setparms\@psfiga,}}
\newif\ifno@bb
\newif\ifnot@eof
\newread\ps@stream
\def\bb@missing{
	\if@verbose{
		\typeout{psfig: searching \@p@sfile \space  for bounding box}
	}\fi
	\openin\ps@stream=\@p@sfile
	\no@bbtrue
	\not@eoftrue
	\catcode`\%=12
	\loop
		\read\ps@stream to \line@in
		\global\toks200=\expandafter{\line@in}
		\ifeof\ps@stream \not@eoffalse \fi
		\@bbtest{\toks200}
		\if@bbmatch\not@eoffalse\expandafter\bb@cull\the\toks200\fi
	\ifnot@eof \repeat
	\catcode`\%=14
}	
\catcode`\%=12
\newif\if@bbmatch
\def\@bbtest#1{\expandafter\@a@\the#1
\long\def\@a@#1
\long\def\bb@cull#1 #2 #3 #4 #5 {
	\dimen100=#2 bp\edef\@p@sbbllx{\number\dimen100}
	\dimen100=#3 bp\edef\@p@sbblly{\number\dimen100}
	\dimen100=#4 bp\edef\@p@sbburx{\number\dimen100}
	\dimen100=#5 bp\edef\@p@sbbury{\number\dimen100}
	\no@bbfalse
}
\catcode`\%=14
\def\compute@bb{
		\no@bbfalse
		\if@bbllx \else \no@bbtrue \fi
		\if@bblly \else \no@bbtrue \fi
		\if@bburx \else \no@bbtrue \fi
		\if@bbury \else \no@bbtrue \fi
		\ifno@bb \bb@missing \fi
		\ifno@bb \typeout{FATAL ERROR: no bb supplied or found}
			\no-bb-error
		\fi
		\count203=\@p@sbburx
		\count204=\@p@sbbury
		\advance\count203 by -\@p@sbbllx
		\advance\count204 by -\@p@sbblly
		\edef\@bbw{\number\count203}
		\edef\@bbh{\number\count204}
}
\def\in@hundreds#1#2#3{\count240=#2 \count241=#3
		     \count100=\count240	
		     \divide\count100 by \count241
		     \count101=\count100
		     \multiply\count101 by \count241
		     \advance\count240 by -\count101
		     \multiply\count240 by 10
		     \count101=\count240	
		     \divide\count101 by \count241
		     \count102=\count101
		     \multiply\count102 by \count241
		     \advance\count240 by -\count102
		     \multiply\count240 by 10
		     \count102=\count240	
		     \divide\count102 by \count241
		     \count200=#1\count205=0
		     \count201=\count200
			\multiply\count201 by \count100
		 	\advance\count205 by \count201
		     \count201=\count200
			\divide\count201 by 10
			\multiply\count201 by \count101
			\advance\count205 by \count201
		     \count201=\count200
			\divide\count201 by 100
			\multiply\count201 by \count102
			\advance\count205 by \count201
		     \edef\@result{\number\count205}
}
\def\compute@wfromh{
		\in@hundreds{\@p@sheight}{\@bbw}{\@bbh}
		\edef\@p@swidth{\@result}
}
\def\compute@hfromw{
		\in@hundreds{\@p@swidth}{\@bbh}{\@bbw}
		\edef\@p@sheight{\@result}
}
\def\compute@wfroms{
		\in@hundreds{\@p@sscale}{\@bbw}{100}
		\edef\@p@swidth{\@result}
}
\def\compute@hfroms{
		\in@hundreds{\@p@sscale}{\@bbh}{100}
		\edef\@p@sheight{\@result}
}
\def\compute@handw{
		\if@scale
			\compute@wfroms
			\compute@hfroms
		\else
			\if@height 
				\if@width
				\else
					\compute@wfromh
				\fi	
			\else 
				\if@width
					\compute@hfromw
				\else
					\edef\@p@sheight{\@bbh}
					\edef\@p@swidth{\@bbw}
				\fi
			\fi
		\fi
}
\def\compute@resv{
		\if@rheight \else \edef\@p@srheight{\@p@sheight} \fi
		\if@rwidth \else \edef\@p@srwidth{\@p@swidth} \fi
}
\def\compute@sizes{
	\compute@bb
	\compute@handw
	\compute@resv
}
\def\psfig#1{\vbox {
	\ps@init@parms
	\parse@ps@parms{#1}
	\compute@sizes
	\ifnum\@p@scost<\@psdraft{
		\if@verbose{
			\typeout{psfig: including \@p@sfile \space }
		}\fi
		\special{ps::[begin] 	\@p@swidth \space \@p@sheight \space
				\@p@sbbllx \space \@p@sbblly \space
				\@p@sbburx \space \@p@sbbury \space
				startTexFig \space }
		\if@clip{
			\if@verbose{
				\typeout{(clip)}
			}\fi
			\special{ps:: doclip \space }
		}\fi
		\if@prologfile
		    \special{ps: plotfile \@prologfileval \space } \fi
		\special{ps: plotfile \@p@sfile \space }
		\if@postlogfile
		    \special{ps: plotfile \@postlogfileval \space } \fi
		\special{ps::[end] endTexFig \space }
		\vbox to \@p@srheight true sp{
			\hbox to \@p@srwidth true sp{
				\hss
			}
		\vss
		}
	}\else{
		\vbox to \@p@srheight true sp{
		\vss
			\hbox to \@p@srwidth true sp{
				\hss
				\if@verbose{
					\@p@sfile
				}\fi
				\hss
			}
		\vss
		}
	}\fi
}}
\def\psglobal{\typeout{psfig: PSGLOBAL is OBSOLETE; use psprint -m instead}}
\catcode`\@=12\relax